\documentclass[letterpaper,twocolumn,10pt]{article}
\usepackage{usenix2019_v3}

\usepackage{tikz}
\usepackage{amsmath}
\usepackage{array}
\usepackage{booktabs}
\usepackage{multirow}
\usepackage{tabularx}
\usepackage{adjustbox}
\usepackage{makecell}
\usepackage{caption}
\usepackage{xspace}
\usepackage{pifont}
\usepackage{amssymb}
\usepackage{tablefootnote}
\usepackage{filecontents}
\usepackage{threeparttable}
\usepackage{subfigure}
\usepackage{xurl}
\usepackage{colortbl} 
\usepackage{xcolor}

\usepackage{algorithm}
\usepackage{algpseudocode}
\usepackage{amsmath} 

\usepackage[available, functional, reproduced]{usenixbadges}

\newcommand{\UAPName}{LS-TUAP\xspace}
\newcommand{\UAPSystemName}{AudioShield\xspace}
\newcommand{\ie}{{\textit{i.e.}},\xspace}
\newcommand{\eg}{{\textit{e.g.}},\xspace}

\newcommand{\etal}{{\textit{et al.}}}

\newcommand{\one}{({\em i}\/)}
\newcommand{\two}{({\em ii}\/)}
\newcommand{\three}{({\em iii}\/)}

\newtheorem{theorem}{Theorem}

\newcommand{\cmark}{\textcolor{blue}{\ding{51}}}
\newcommand{\xmark}{\textcolor{red}{\ding{55}}} 
\newcommand{\solidstar}{\ding{72}}

\newcommand{\solidcircle}{\ding{108}}
\newcommand{\hollowcircle}{\ding{109}}
\newcommand{\solidsquare}{\ding{110}}
\newcommand{\hollowsquare}{\ding{111}}
\newcommand{\solidtriangle}{\ding{115}}

\newcommand{\hollowdiamond}{\ding{118}}

\begin{document}
%-------------------------------------------------------------------------------
\captionsetup{skip=10pt} 

\date{}

\title{\Large \bf 
Whispering Under the Eaves: Protecting User Privacy Against Commercial and LLM-powered Automatic Speech Recognition Systems
}

\rm 
\author{Weifei Jin$^\dagger$, Yuxin Cao$^\ddagger$, Junjie Su$^\dagger$, Derui Wang$^\S$, Yedi Zhang$^\ddagger$, Minhui Xue$^\S$, \\ 
Jie Hao$^{\dagger}$, Jin Song Dong$^\ddagger$, Yixian Yang$^\dagger$\\
\small $^\dagger$Beijing University of Posts and Telecommunications \small \hspace{0.5cm}  $^\ddagger$ National University of Singapore \small \hspace{0.5cm} $^\S$ CSIRO's Data61 \\
}

\maketitle

%-------------------------------------------------------------------------------
\begin{abstract}
The widespread application of automatic speech recognition (ASR) supports large-scale voice surveillance, raising concerns about privacy among users. In this paper, we concentrate on using adversarial examples to mitigate unauthorized disclosure of speech privacy thwarted by potential eavesdroppers in speech communications. While audio adversarial examples have demonstrated the capability to mislead ASR models or evade ASR surveillance, they are typically constructed through time-intensive offline optimization, restricting their practicality in real-time voice communication. Recent work overcame this limitation by generating universal adversarial perturbations (UAPs) and enhancing their transferability for black-box scenarios. However, they introduced excessive noise that significantly degrades audio quality and affects human perception, thereby limiting their effectiveness in practical scenarios. To address this limitation and protect live users' speech against ASR systems, we propose a novel framework, \UAPSystemName. Central to this framework is the concept of \textbf{T}ransferable \textbf{U}niversal \textbf{A}dversarial \textbf{P}erturbations in the \textbf{L}atent \textbf{S}pace (\UAPName). By transferring the perturbations to the latent space, the audio quality is preserved to a large extent. Additionally, we propose target feature adaptation to enhance the transferability of UAPs by embedding target text features into the perturbations. Comprehensive evaluation on four commercial ASR APIs (Google, Amazon, iFlytek, and Alibaba), three widely-used voice assistants, two LLM-powered ASR and one NN-based ASR demonstrates the protection superiority of \UAPSystemName over existing competitors, and both objective and subjective evaluations indicate that \UAPSystemName significantly improves the audio quality. Moreover, \UAPSystemName also shows high effectiveness in the real-time end-to-end scenarios, and demonstrates strong resilience against adaptive countermeasures. 
\end{abstract}

%-------------------------------------------------------------------------------
\section{Introduction}

Automatic speech recognition (ASR) systems use deep learning technology to transcribe speech into text, and their high accuracy has led to widespread application across various fields~\cite{wang2019overview,mehrish2023review,poushneh2021humanizing,yan2022survey}. For example, in the financial services sector, voice surveillance is used to improve the reliability and efficiency of compliance management~\cite{financial}. However, indiscriminate voice surveillance can raise security concerns, particularly in the context of global government agencies extensively monitoring personal phone calls and internet data~\cite{surveillance}. This large-scale voice surveillance has sparked public concerns about personal privacy breaches and eroded trust in governments. Therefore, protecting public privacy and avoiding voice surveillance have become increasingly necessary.

\begin{figure}[t]
    \centering
    \includegraphics[width=0.4\textwidth]{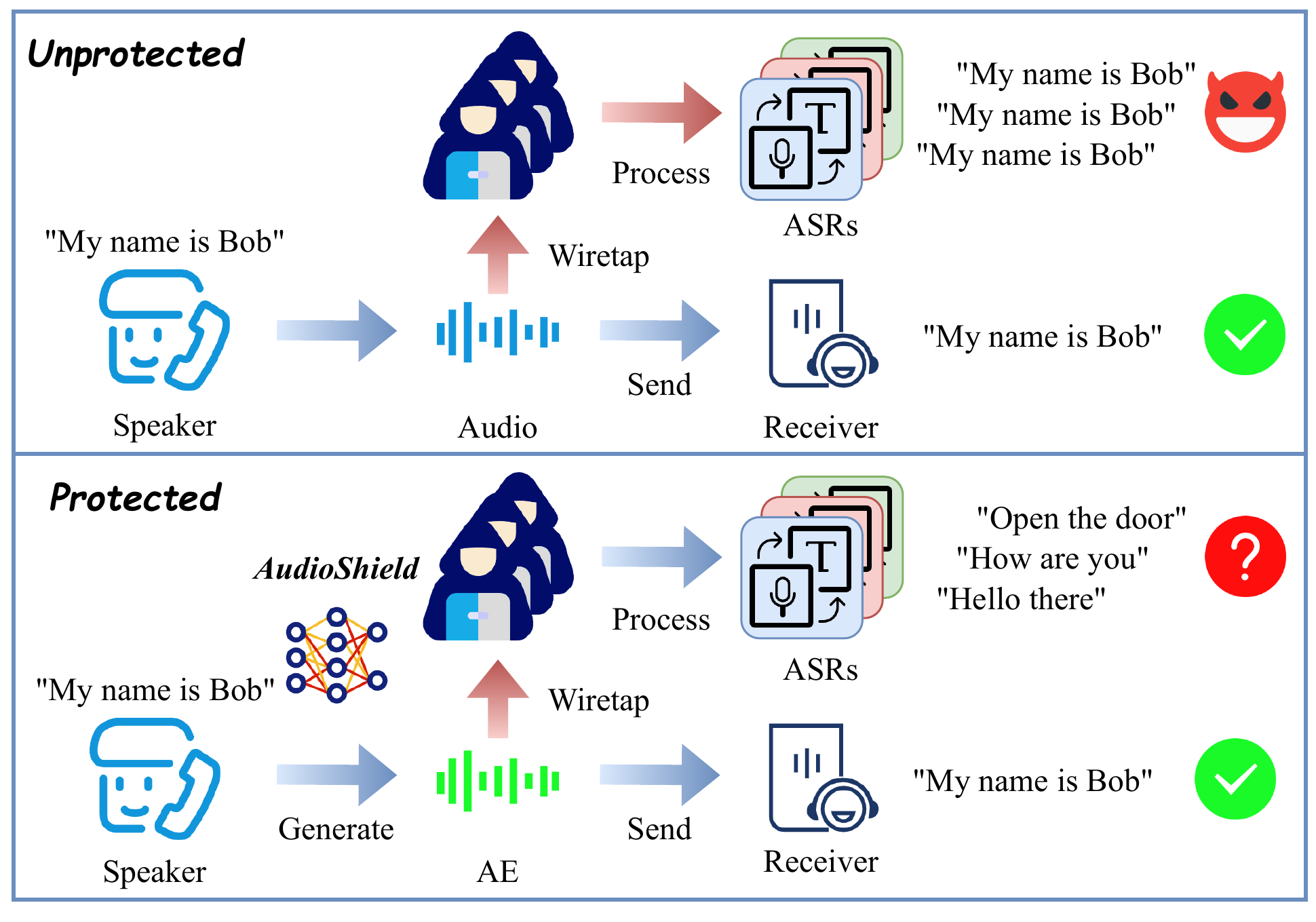}
    \caption{Overview of large-scale speech communication surveillance scenarios without/with \UAPSystemName.}
    \label{fig:lsuap_scenario}
    \vspace{-3mm}
\end{figure}

\begin{table*}[t]
\centering
\caption{Comparison of existing research focusing on audio adversarial examples.}
\label{tab:comparison_methods}
\resizebox{0.8\linewidth}{!}{
\begin{tabular}{c|c|c|c|c|c|c}
\Xhline{1px}
\textbf{Method} & \textbf{Knowledge} & \textbf{Transferability} & \textbf{Universality} & \textbf{Unrestriction}\footnotemark[1] & \textbf{Target ASR} & \textbf{Applicability} \\ \Xhline{1px}
Carlini \etal~\cite{carlini2018audio} & \hollowsquare & \xmark & \xmark & \xmark & \solidtriangle & \hollowcircle\\ \hline
Neekhara \etal~\cite{neekhara2019universal} & \hollowsquare & \xmark & \cmark & \xmark & \solidtriangle & \hollowcircle\\ \hline
Zong \etal~\cite{zong2021targeted} & \hollowsquare & \xmark & \cmark & \xmark & \solidtriangle & \hollowcircle\\ \hline
SSA~\cite{qu2022synthesising} & \hollowsquare & \xmark & \xmark & \cmark & \solidtriangle & \hollowcircle \\ \hline
Devil's Whisper~\cite{chen2020devil} & \solidsquare & \xmark & \xmark & \xmark &   \hollowdiamond & \solidcircle\\ \hline
KENKU~\cite{wu2023kenku} & \solidsquare & \xmark & \xmark & \xmark &  \hollowdiamond & \solidcircle\\ \hline
SMACK~\cite{yu2023smack} & \solidsquare & \xmark & \xmark & \cmark & \solidtriangle \hollowdiamond & \solidcircle\\ \hline
Transaudio~\cite{qi2023transaudio} & \solidsquare & \cmark & \xmark & \xmark & \solidtriangle \hollowdiamond  & \hollowcircle\\  \hline
ZQ-Attack~\cite{fang2024zero} & \solidsquare & \cmark & \xmark & \xmark & \solidtriangle \hollowdiamond& \solidcircle\\ \hline
AdvDDoS~\cite{ge2023advddos} & \solidsquare & \cmark & \cmark & \xmark & \solidtriangle \hollowdiamond & \solidcircle\\ \hline
\UAPSystemName & \solidsquare & \cmark & \cmark & \cmark & \solidtriangle \hollowdiamond\solidstar & \solidcircle\\ \Xhline{1px}
\end{tabular}
}
    \\
    \parbox{0.88\textwidth}{
    \footnotesize
    \hollowsquare: represents white-box settings; 
    \solidsquare: represents black-box settings.
    
    \solidtriangle: represents open-sourced traditional NN-based ASR models; \hollowdiamond: represents commercial ASR models; \solidstar: represents LLM-powered ASR models.

    \hollowcircle: represents over-the-line (digital) settings; 
    \solidcircle: represents both over-the-line (digital) and over-the-air (physical) settings.

    \footnotemark[1]: ``restriction'' means that the perturbation between the adversarial example and the original audio is constrained by traditional $\ell_p$-norms; ``unrestriction'' represents that the adversarial example can have significant variations compared to the original audio. 
    }
    \vspace{-3mm}
\end{table*}

As shown in Figure~\ref{fig:lsuap_scenario}, the large-scale speech communication surveillance typically includes three parties, speaker, receiver, and eavesdropper. While the speaker conveys speech to the receiver, the eavesdropper can seize the opportunity to intercept large amounts of user speech data and use ASR to convert them into texts for quick extraction of key information. More seriously, due to the lack of any protection for speech, the speaker and receiver may never know and never come to know that their conversations are being monitored. In summary, such unprotected conversations provide the possibility of privacy content leakage to third parties.

Adversarial perturbations against ASR systems offer a potential avenue for evading voice surveillance by adding subtle perturbations to the original audio data, causing the target ASR model to produce erroneous transcriptions.
Specifically, the protection service provider offers protection services on the user side that converts the user's speech into adversarial examples, preventing the ASR from correctly recognizing the speech and thereby protecting the privacy information contained in the user's speech.

Researchers in this field have proposed various adversarial examples against ASR systems~\cite{liu2022evil, guo2022specpatch, chiquier2022real, yu2023smack,cheng2024alif}, but they are never used to protect the users' speech privacy due to weaknesses of cost-consuming, low-transferability or low-quality.
A white-box setting implies that the structure and parameters of the target model cannot be accessed, and therefore it is not suitable for the scenarios described above, since no internal knowledge of the practical system can be obtained. In contrast, black-box settings do not rely on knowledge of the target model, which increases their practicality. Black-box adversarial examples are mainly divided into two categories: query-based and transfer-based, with most adversarial examples against ASR systems currently being query-based~\cite{bhanushali2024adversarial}. 
Table~\ref{tab:comparison_methods} provides a comprehensive comparison of existing research in this field.
However, in the context of evading voice communication surveillance, these methods often suffer from three significant drawbacks. \one~\textbf{Cost-consuming}: query-based perturbations require sufficient time and queries for constructing adversarial examples, which is impractical given the real-time requirements of voice communication. An intuitive solution to solve this problem is to train universal perturbations in advance, and then insert them to real-time audio.
\two~\textbf{Low-transferability}: query-based perturbations are designed on a single target model, which does not align with the reality that users do not know the specific ASR model used by the surveillant in practical scenarios. Therefore, such adversarial perturbations lack transferability across different, especially unseen models. By leveraging the transferability of adversarial examples, transferable perturbations, which are first crafted on a local surrogate model, and then transferred to the target model, can partially solve the above two limitations. \three~\textbf{Low-quality}: different from images in which the imperceptibility of adversarial perturbations can be bounded within the $\ell_p$-norm ball, the perturbation restriction of audio data is much more difficult to define. Some existing work uses decibel distortion, Signal-to-Noise Ratio, or Perceptual Evaluation of Speech Quality~\cite{rix2001perceptual} as restriction metrics to maintain the quality. However, the adversarial audios of most existing work~\cite{neekhara2019universal,zong2021targeted,ge2023advddos} still include harsh noise which makes it hard for humans to figure out the audio contents.
Therefore, how to handle all these drawbacks to achieve both adversariality and high audio quality remains unsolved.

To overcome these limitations, we propose \UAPSystemName by introducing Transferable Universal Adversarial Perturbations in the Latent Space (\UAPName), which shifts the perturbations to the latent space, thereby avoiding the introduction of noise into the original audio data space and preserving audio quality. Specifically, we utilize a variational autoencoder (VAE) architecture, where the input audio is encoded into latent space by the encoder. Perturbations are then added in the latent space, and the perturbed latent codes are passed through the decoder to synthesize the audio back. Our method aligns with the requirements for evading voice communication surveillance in three key aspects: \one~\textbf{Real-time Requirement}: we train universal adversarial perturbations in the latent space, making them effective across any audio input. This eliminates the need to iteratively generate perturbations for each specific audio, thus meeting the real-time requirement. \two~\textbf{Model-agnostic Requirement}: we propose a target feature adaptation process to enable \UAPName to learn the robust features of the target text, enhancing its transferability. This makes \UAPSystemName effective against unseen models, thereby satisfying the model-agnostic requirement. \three~\textbf{High-quality Requirement}: our perturbations are applied in the latent space, rather than the audio input. We also find a $r$-robustness probability bound for the output of the decoder,
indicating that \UAPName can 
ensure the preservation of audio semantics and quality in voice communication.

To validate the effectiveness of our method, we conduct extensive experiments on four commercial ASR APIs, two LLM-powered ASR, one NN-based ASR, and three voice assistant devices. Through comparisons with competitors, the superiority of our method is clearly demonstrated. Specifically, in the over-the-line setting, \UAPSystemName achieves a protection success rate of over 75\% on four commercial ASR APIs, two LLM-powered models and one NN-based model, 
surpassing the most advanced competitor (AdvDDoS) by 27.88\%, 27.44\%, 5.5\%, and 17.29\% on Google, Amazon, iFlytek, and Alibaba, respectively. To further verify audio quality, we perform both objective and subjective evaluations, where the adversarial examples generated by our method significantly outperform those by existing methods. Furthermore, \UAPSystemName achieves an average of 87.5\% and 69\% protection success rate in the real-time end-to-end evaluation and over-the-air robustness evaluation, respectively. 
\UAPSystemName also exhibits stronger resilience to adaptive countermeasures compared to competitors. 
The source code and audio demos are available at~\url{https://github.com/WeifeiJin/AudioShield}.

In summary, our contributions are as follows:
\begin{itemize}
    \item We propose a novel framework, AudioShield, designed to protect live users’ speech against ASR models in large-scale voice surveillance. The core of AudioShield involves the introduction of LS-TUAP, which achieves high universality, transferability, and audio quality.
    \item We introduce a target feature adaptation process that optimizes the similarity loss in latent space, enabling the perturbation to learn robust features of the target text, thereby enhancing the transferability of LS-TUAP.
    \item We conduct extensive evaluations on ten ASR models, including four commercial ASR APIs, two LLM-powered models, one NN-based model, and three voice assistants. The experimental results demonstrate the superiority of \UAPSystemName, surpassing competitors in both protection performance and audio quality. 
\end{itemize}

\section{Background and Related Work}

\subsection{Automatic Speech Recognition}
\begin{figure}[t]
    \centering\includegraphics[width=0.9\linewidth]{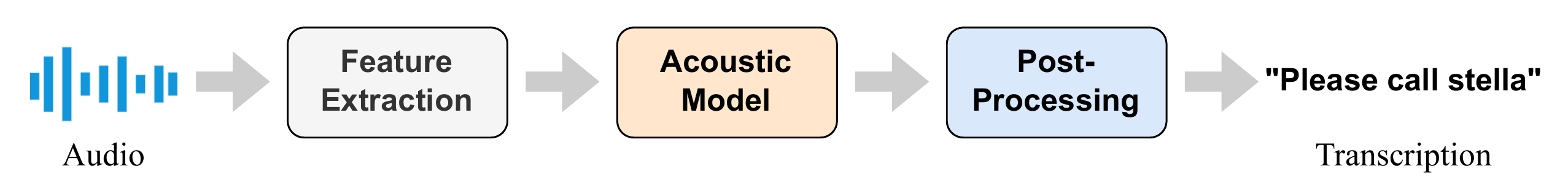}
    \caption{The architecture of a typical ASR system.}
    \label{fig:asr_workflow}
    \vspace{-3mm}
\end{figure}

An ASR system transforms audio into text. Given an input audio $x$, the ASR model $f(\cdot)$ generates the transcription $y$ such that $y = f(x)$. The typical architecture of the ASR system is illustrated in Figure~\ref{fig:asr_workflow}, which consists of three components: feature extraction, acoustic model, and post-processing.
The acoustic characteristics are first captured by the feature extraction process. Traditional systems use time-frequency transformations to convert audio into a frequency-domain spectrogram~\cite{gupta2016lpc,al2017capacity}, while modern systems often use the raw spectrogram directly~\cite{amodei2016deep}.
Then, the acoustic model maps the extracted features or the raw spectrogram to an intermediate representation using Gaussian Mixture Model-Hidden Markov Models~\cite{gales2008application} or deep neural network (DNNs)~\cite{li2019jasper, hori2018end, gulati2020conformer}.
Finally, the output of the acoustic model, \ie tokens, is converted into texts in the post-processing step.

Audio adversarial examples can mislead the audio system into producing incorrect outputs or performing incorrect behaviors. Existing studies target different tasks including speaker recognition~\cite{chen2021real,chen2023qfa2sr}, speech command classification~\cite{zhang2021commandergabble}, sound event classification~\cite{tripathi2022adv} and speech recognition~\cite{yuan2018commandersong}. In this paper, we concentrate on deceiving the ASR system into producing incorrect transcriptions to protect user privacy from content leakage.
Adversarial examples on ASR systems can be crafted in targeted and untargeted settings. In the untargeted setting, the ASR system is misled into producing any transcription other than the correct one, represented as $f(x') \neq y$, where $x'$ denotes the adversarial example. Conversely, in the targeted setting, the ASR system is misled to produce a specific incorrect transcription $t \neq f(x)$, formulated as $f(x') = t$. The adversarial example is also subject to $\text{d}(x, x') \le \epsilon$, where $\text{d}(x, x')$ measures the distance between $x$ and $x'$. $\ell_p$ norm is commonly used to calculate the distance, and $\epsilon$ constrains the perturbation magnitude.

\subsection{Related Work}

Although there are various types of taxonomy for research on audio adversarial examples (as shown in Table~\ref{tab:comparison_methods}), we introduce these studies by roughly categorizing the perturbations into universal and transferable audio adversarial perturbations, which is the most relevant classification method for this paper.

\noindent\textbf{Universal audio adversarial perturbations.}
Universal perturbations mislead the target model across a wide range of inputs. They are trained offline and then applied to online inputs, making them suitable for real-time applications like audio and video streaming.
Apart from work that proposes UAPs in audio classification tasks~\cite{li2020advpulse,xie2021enabling}, 
Neekhara \etal~\cite{neekhara2019universal} first proposed untargeted UAPs to achieve input-agnostic manipulations in ASR tasks, followed by Zong \etal~\cite{zong2021targeted} who introduced a two-stage method to generate targeted UAPs.
However, these works are still based on white-box settings, which overestimate the manipulator's capability. 
Recently, AdvDDoS~\cite{ge2023advddos} was proposed to generate transferable universal adversarial perturbations, also using a two-stage method and leveraging mel-frequency cepstral coefficients (MFCC) feature inversion to enhance the transferability. 
Although these methods employ two-stage algorithms to minimize perturbation size as much as possible, sufficiently large perturbations are still necessary to achieve universality, which results in excessive noise in the generated adversarial examples, thereby degrading audio quality.

\noindent\textbf{Transferable audio adversarial perturbations.}
In order to reduce query budgets or even achieve query-free manipulations in the black-box setting, transferable adversarial perturbations are proposed by utilizing the transferability of adversarial examples across different models. Adversarial examples are first trained on a known surrogate model and then transferred to other unseen models. However, research on transferable perturbations is still in their infancy for ASR tasks. 
To mitigate the overfitting problem of optimizing adversarial examples on the surrogate model, Qi \etal~\cite{qi2023transaudio} proposed Transaudio, which is a contextualized manipulation including various adversarial behaviors. 
However, the performance of universality and long audios was not verified. Ge \etal~\cite{ge2023advddos} proposed AdvDDoS, which both considers transferablility and universality when generating adversarial examples, but largely decreases the audio quality which can make adversarial audios detectable and conspicuous. Recently, Fang \etal~\cite{fang2024zero} proposed ZQ-Attack, which is an ensemble method that uses different types of surrogate models to enhance the transferablity of the adversarial examples. However, considering various surrogate models during the optimization requires a high computational cost, and the audio quality of generated perturbations is not well preserved. Therefore, in this paper, to improve the practicality of audio adversarial examples, we consider enhancing both universality and transferability while maintaining the audio quality.

\section{Threat Model}
\begin{figure}[t]
    \centering\includegraphics[width=0.9\linewidth]{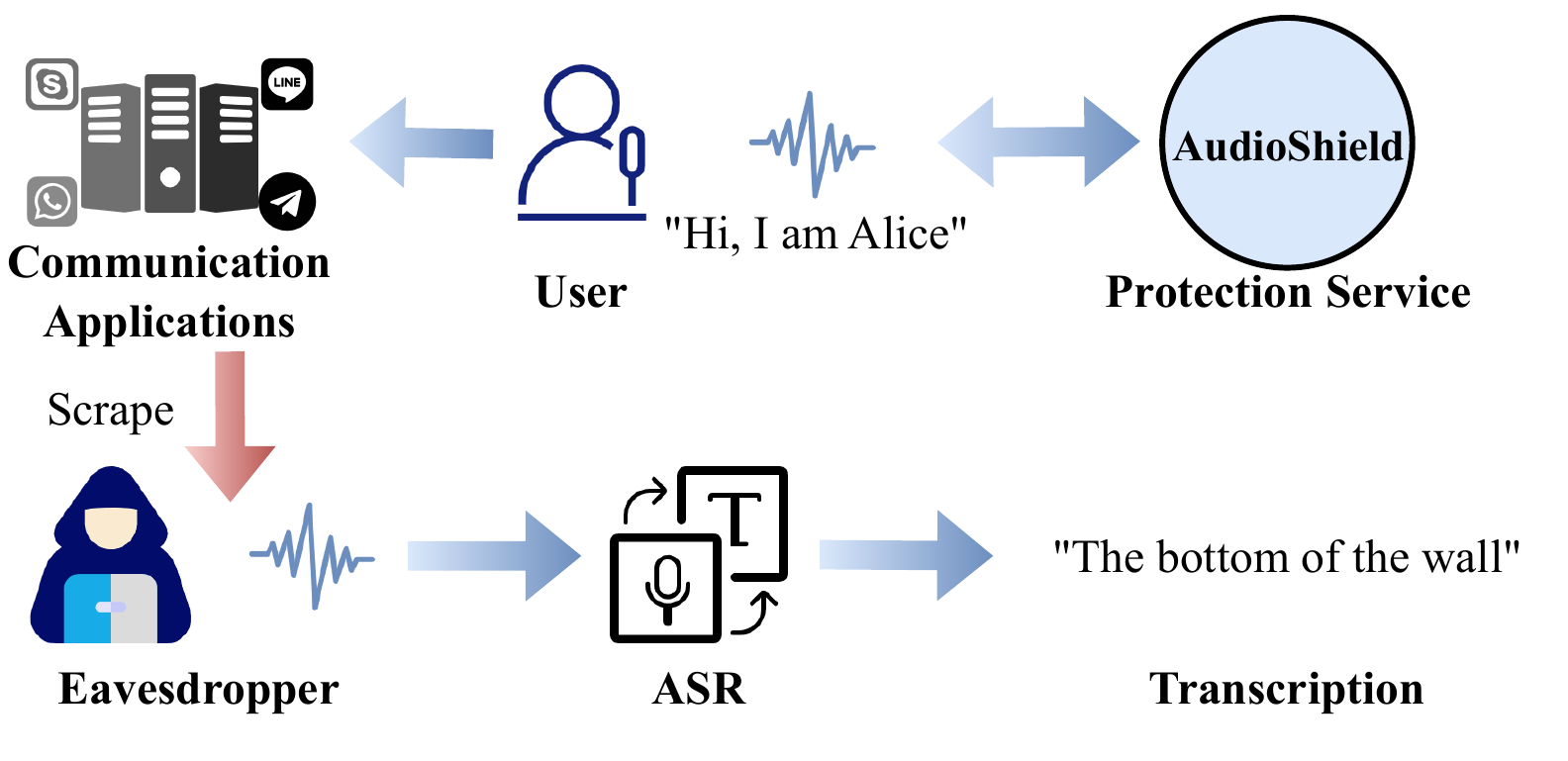}
    \vspace{-2mm}
    \caption{Protection scenario of \UAPSystemName.}
    \label{fig:scenario}
    \vspace{-3mm}
\end{figure}

\noindent\textbf{Protection Scenario.} In the scenario we assume, as shown in Figure~\ref{fig:scenario}, there are three parties involved: the protection service provider (such as \UAPSystemName), the user, and the eavesdropper. The user's speech communication data may be intercepted by potential eavesdroppers in the real world without authorization. Eavesdroppers use ASR systems to conduct large-scale surveillance on the speech communications of many users. \UAPSystemName packages well-trained perturbations into a program or software that can run in the background on local devices or be provided as a cloud service. \UAPSystemName offers a service that protects users' speech privacy by receiving their speech input, applying protection using \UAPName, and then outputting the protected audio through a virtual microphone to downstream speech communication software. By using \UAPSystemName, users can protect the privacy of their speech data. The eavesdropper, can directly access large amounts of user communication data from government agencies or large companies' servers, as reported in~\cite{surveillance}.

\noindent\textbf{System Goal.} \UAPSystemName converts each normal speech input into an adversarial example, making its semantic content transcribed incorrectly by ASR systems. At the same time, the adversarial perturbations generated by \UAPSystemName must be universal to meet the real-time requirements of speech communication scenarios. For any speech input, only a single inference process is needed, ensuring that the latency remains within an acceptable range. \UAPSystemName can function as a background program or cloud service, receiving the user's speech input, converting it into adversarial examples, and then outputting it to communication software or channels, so that eavesdroppers only receive the adversarial examples. Furthermore, the adversarial examples generated by \UAPSystemName must maintain a certain level of quality, allowing the human recipient in the communication to still understand its content.

\noindent\textbf{Knowledge Assumption.} In our assumed scenario, both the protection service provider (\UAPSystemName) and the user operate under a completely black-box setting, meaning they have no access to the architecture, parameters, or any output of the target model. This is because no information about the potential eavesdropper is available, and thus, the specific ASR model being used by the eavesdropper is unknown. This requires the adversarial examples generated by \UAPSystemName to have high transferability, ensuring strong protection across various ASR models that the eavesdropper might use. In other words, different ASR systems should not be able to accurately recognize the semantic content of the speech. To achieve this, we train the adversarial perturbations using a surrogate model locally and then transfer them to black-box ASR models in an untargeted setting.

\section{Design of \UAPSystemName}

\subsection{Problem Formulation}
Our goal is to generate transferable universal adversarial perturbation that is effective on any \textit{unseen} audio and any \textit{unseen} black-box ASR model, \ie in the untargeted setting. To achieve this goal, perturbations are first crafted in the targeted setting on a local surrogate model. Since the surrogate model is accessible, we can obtain all information about the model. Therefore, the objective function can be expressed as follows.
\begin{equation}
    \min_{\delta_x} \mathop{\mathbb{E}}_{x \sim \mathcal{X}} [\mathcal{L}_{ASR}(f(x'), t)],
\label{eq:basic}
\end{equation}
where $\mathcal{X}$ denotes the audio dataset, $t$ represents the target text, $\delta_x$ stands for the perturbation that we want to obtain. The ASR loss $\mathcal{L}_{ASR}$ is defined as connectionist temporal classification (CTC) loss or cross-entropy loss, according to the architecture of the surrogate model. As mentioned previously, in traditional methods that add perturbations to the input audio, $x'$ satisfies $x' = x + \delta_x$, and $\text{d}(x, x') \le \epsilon $. However, in our work, perturbations are added in the latent space, which will be introduced later. After the UAPs are obtained, they are added to real-time audios and the generated audios are input to unseen ASR models for further testing.

\subsection{Overview of \UAPSystemName}
\begin{figure}[t]
\centering
\includegraphics[width=0.98\linewidth]{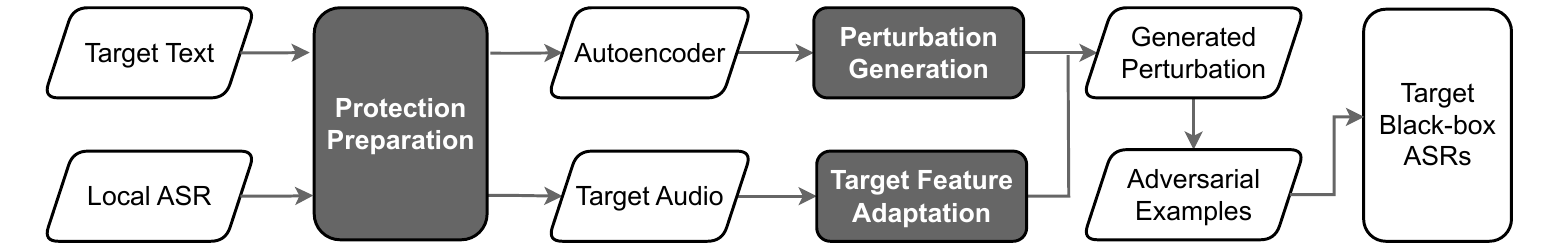}
\caption{Workflow of \UAPSystemName. The three main steps of the whole process are: protection preparation, perturbation generation and target feature adaptation.}
\label{fig:workflow}
\vspace{-3mm}
\end{figure}

The core idea of \UAPSystemName is to optimize a transferable universal adversarial perturbation in the latent space. Specifically, as illustrated in Figure~\ref{fig:workflow}, the optimization process is primarily divided into three steps: protection preparation, perturbation generation, and target feature adaptation. The pseudocode of \UAPSystemName can be found in Algorithm~\ref{alg:audioshield}.

\begin{algorithm}[!t]
\caption{\UAPSystemName}
\label{alg:audioshield}
\renewcommand{\algorithmicrequire}{\textbf{Input:}}
\renewcommand{\algorithmicensure}{\textbf{Output:}}
\begin{algorithmic}[1]
\Require Encoder $\mathcal{E}$, decoder $\mathcal{D}$, ASR model $f$, TTS model $g$, target text $t$, training dataset $\mathcal{X}$, standard deviation $\sigma$, maximum epoch number $MaxEpoch$, maximum iteration for each batch $MaxIter$, target audio number $n$, learning rate $\alpha$, perturbation threshold $\tau$, decay rate $s$, ASR loss $\mathcal{L}_{ASR}$, cosine similarity loss $\mathcal{L}_{Sim}$, weighting factor $\lambda$.
\Ensure Generated \UAPName $\delta$.
\State Initialization: $MinL \gets +\infty$, $x_t \gets 0$, $\delta \gets 0$
\For{$i \leftarrow 1$ \textbf{to} $n$}
    \State Randomly generate audio $x_{t_i}\gets g(t)$
    \State $z_i \gets \mathcal{E}(x_{t_i})$, $w \gets 1$
    \While{$f(\mathcal{D}(w\cdot z_i)) == t$}
        \State $w \gets w\cdot s$ 
    \EndWhile
    \State $z_i\gets w\cdot z_i$, $x_{t_i}\gets \mathcal{D}(z_i)$
    \State $l_i \gets \mathcal{L}_{ASR}(f(x_{t_i}), t)$
    \If{$l_i < MinL$}
        \State $x_t \gets x_{t_i}$, $\delta \gets z_i$, $MinL \gets l_i$
    \EndIf
\EndFor

\For{$i \leftarrow 1$ \textbf{to} $MaxEpoch$}
    \State Randomly sample a batch audio $x$ from $\mathcal{X}$
    \State $z \gets \mathcal{E}(x)$
    \For{$j \leftarrow 1$ \textbf{to} $MaxIter$}
        \State Sample $p \sim \mathcal{N}(0, \sigma^2)$
        \State $x' \gets \mathcal{D}(z + \delta + p)$
        \State Calculate $\mathcal{L}_{total} \gets \mathcal{L}_{ASR}(f(x',t))+\lambda \cdot \mathcal{L}_{Sim}(\delta, x_t)$
        \State $\delta \gets \text{clip}(\delta - \alpha \cdot \text{sign}(\nabla_{\delta} \mathcal{L}_{total}), -\tau, \tau)$
    \EndFor
\EndFor
\State \Return $\delta$
\end{algorithmic}
\end{algorithm}

\noindent\textbf{Protection Preparation.} The main objective of this step is to select an appropriate autoencoder and a suitable target audio. For autoencoder selection, we establish three principles to guide the selection process. We then use a heuristic search algorithm to find a suitable target audio and the scaling factor, which serves as the desiderata for subsequent steps.

\begin{figure}[t]
\centering
\includegraphics[width=0.9\linewidth]{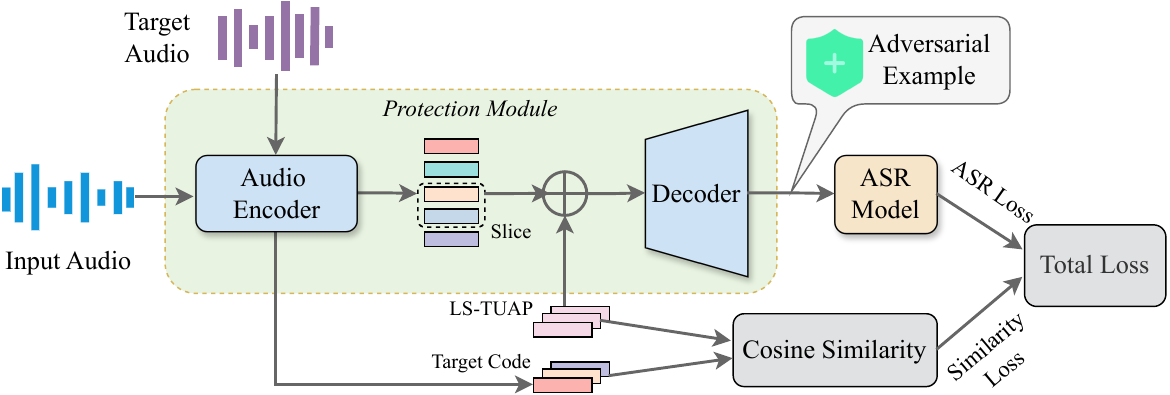}
\caption{Illustration of perturbation generation and target feature adaptation.}
\label{fig:illustration_s2_s3}
\vspace{-2mm}
\end{figure}

\noindent\textbf{Perturbation Generation.} This is the core step of the entire optimization process. As shown in Figure~\ref{fig:illustration_s2_s3}, we utilize an autoencoder architecture as the protection module where the audio is first input to an encoder to obtain its corresponding latent code. The adversarial perturbation, \UAPName, is then added to the latent code, and the perturbed latent code is sent to a decoder to synthesize the adversarial example. 

\noindent\textbf{Target Feature Adaptation.} To enhance the transferability of \UAPName, we propose target feature adaptation. Also as shown in Figure~\ref{fig:illustration_s2_s3}, the target audio generated from target text is fed into the encoder to obtain its latent code. By minimizing the cosine similarity between \UAPName and the target audio's latent code, the perturbation learns the latent features of the target text.

\noindent\textbf{Over-the-air Robustness.} 
For the physical scenario, we employ room impulse response (RIR) to model the transfer function between the sound source and the microphone, simulating over-the-air distortions during the perturbation optimization, thereby enhancing its robustness.

\noindent\textbf{Real-time Protection}
Based on the previous steps, we can obtain different UAPs using different target texts. In practical use, each time we obtain the audio in real-time scenarios, we randomly select one UAP from a set of well-generated UAPs, perform inference using the autoencoder we select, and obtain the protected example. Therefore, the latency we introduce is solely the inference time of the autoencoder, which  ensures that the latency remains relatively small. 

\begin{figure}[t]
    \centering\includegraphics[width=0.9\linewidth]{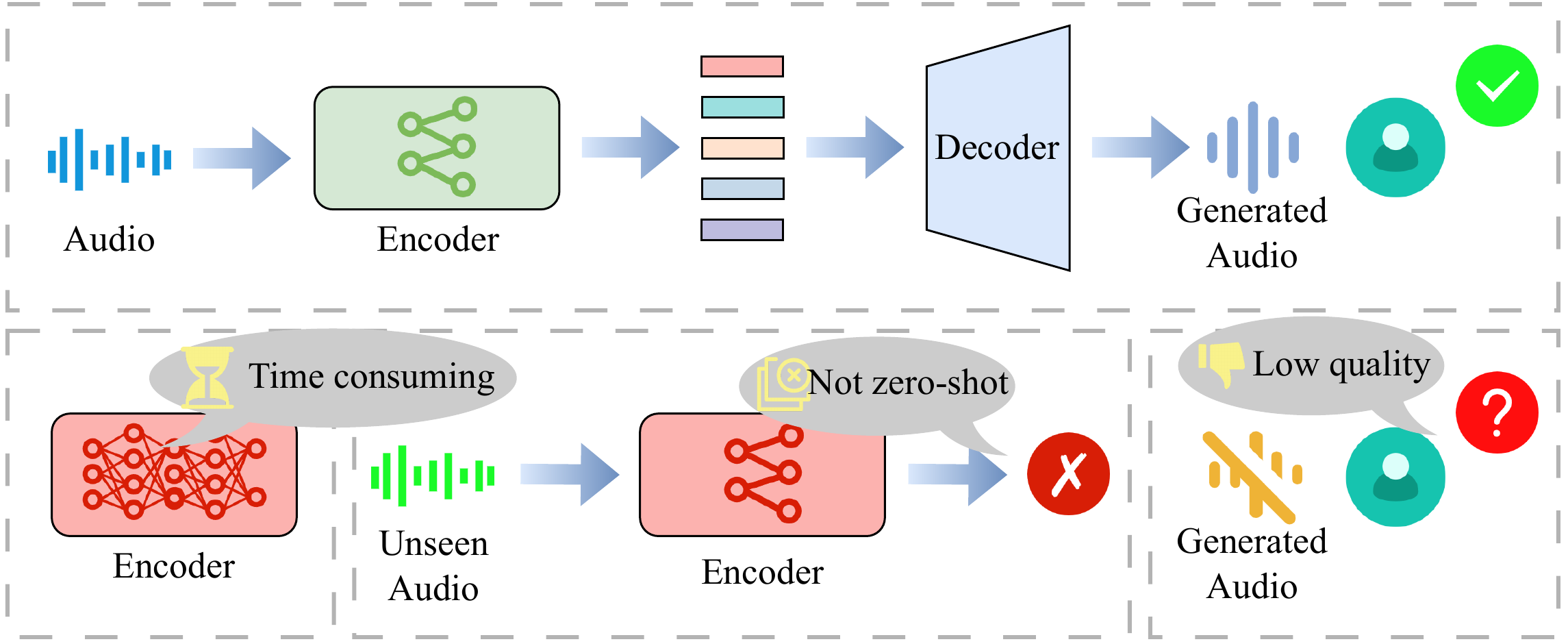}
    \caption{We exclude models based on three criteria: \one~the model has a large parameter size, leading to excessive latency during inference; \two~the model is not zero-shot and thus ineffective on arbitrary audio inputs; \three~the audio generated by the model is low-quality.
    }
    \label{fig:selection_criteria}
    \vspace{-3mm}
\end{figure}

\begin{table}[t]
\centering
\caption{Comparison of different autoencoders.}
\label{tab:autoencoder_selec}
\resizebox{1.0\linewidth}{!}{
\begin{tabular}{c|c|c|c|c}
\Xhline{1px}
\textbf{Model} & \textbf{NISQA} & \textbf{Zero-shot} & \textbf{Param. Size} & \textbf{Infer. Time} \\ 
\Xhline{1px}
AutoVC~\cite{qian2019autovc} & 2.15 & Y & 27.11M & 537ms \\ \hline
SpeechSplit2~\cite{chan2022speechsplit2} & 3.23 & N & 22.71M & 1136ms  \\ \hline
\rowcolor{gray!20} VITS~\cite{kim2021conditional} & 3.84 & Y & 37.86M & 876ms  \\ \hline
HierSpeech++~\cite{lee2023hierspeech++} & 3.33 & Y & 197.99M & 8646ms  \\ 
\Xhline{1px}
\end{tabular}
}
\vspace{-3mm}
\end{table}

\subsection{Protection Preparation}\label{subsec:protection_preparation}

\noindent\textbf{Autoencoder Selection.} 
The efficacy of the audio adversarial examples exhibits a significant correlation with the performance of the VAE utilized. Consequently, the discreet selection of an appropriate autoencoder is of paramount importance.
Based on our protection requirements, we establish the following three principles for autoencoder selection:
\one~\textbf{Efficiency and Compactness}: the model should be streamlined and efficient, with a minimal number of parameters, to ensure that the inference stage does not introduce excessive latency.
\two~\textbf{Zero-shot Capability}: since we aim to generate UAPs that are applicable to any audio, the model must operate in a zero-shot manner.
\three~\textbf{High Audio Quality}: the generated audio must maintain high quality, and the latent code should be able to tolerate a certain degree of perturbation without causing significant degradation in the generated audio.

Based on these three criteria, we evaluate several popular voice conversion models and text-to-speech (TTS) models in the audio domain. 
The results are shown in Table~\ref{tab:autoencoder_selec}, where NISQA~\cite{mittag2021nisqa} is a metric to assess audio quality, which will be introduced in detail in Section~\ref{subsec:experiment_setup}. Considering audio quality, latency, zero-shot capability, and computational resource consumption, we ultimately choose the VITS model~\cite{kim2021conditional}.
Although VITS is primarily a TTS model, it includes a spectrogram encoder component that allows it to accept audio inputs. VITS is a widely recognized model and is often used as the foundational backbone for more advanced TTS models, such as YourTTS~\cite{casanova2022yourtts} and HierSpeech~\cite{lee2022hierspeech}.

\noindent\textbf{Target Audio Selection.}
In the local targeted setting, the perturbations in the latent space requires learning the latent features of the target text to improve its transferability. A better target audio generated from target text makes the perturbations easier to improve the ability of adversarial examples to mislead unknown ASR models and protect user's speech content in the untargeted setting.

To achieve this goal, we propose a heuristic search algorithm to find a suitable target audio and a scaling factor before optimizing the perturbation. Specifically, given a target text $t$, we first use a one-to-many TTS model that allows speaker specification to generate a set of $n$ target audios with different styles and random speakers. Since the VITS model selected in the autoencoder selection step is a TTS model that can achieve one-to-many mapping through adjusting its noise scale parameter, we directly use the VITS model in this step. We then iterate through this set of $n$ target audios. For each target audio $x_{t_i}$, we input it into the audio encoder to obtain the latent code $z_i$. Next, we search for a scaling factor $w$ that minimizes the factor required for the local ASR model to correctly transcribe $\mathcal{D}(w \cdot z_i)$ as the target text $t$, where $\mathcal{D}$ denotes the decoder. We then calculate the corresponding ASR loss. Finally, we select the target audio $x_t$ with the smallest loss value from the $n$ target audios, and the corresponding smallest scaling factor is used to initialize the perturbation $\delta \leftarrow w \cdot \mathcal{E}(x_t)$, where $\mathcal{E}$ denotes the encoder. 

In summary, the heuristic search algorithm identifies the most suitable target audio and the corresponding perturbation initialization scaling factor in a locally optimal sense through a limited search process within a two-tier loop. This approach avoids the brute-force search for the globally optimal target audio and scaling factor in an infinite space and reduces the interference of irrelevant features in the audio during the perturbation optimization process. This approach also reduces the interference of irrelevant features in the audio during the perturbation optimization process. The selection of the target audio can be seen as a coarse-grained optimization of the perturbation, serving as the foundation for fine-grained optimization in subsequent steps.

\subsection{Perturbation Generation}

Given an audio clip $x$ and an adversarial perturbation $\delta$ in the latent space, the adversarial example $x'$ can be expressed as:
\begin{equation}
x' = \mathcal{D}(\mathcal{E}(x) + \delta).
\end{equation}

Meanwhile, we empirically verify that the latent codes in the latent space cannot change arbitrarily, as this would generate audio that sounds natural to humans (more details are provided in Section~\ref{subsec:quality}). To ensure that the distribution of the generated audio is similar to that of the natural audio, we restrict the perturbation within a certain range. Therefore, Equation (\ref{eq:basic}) is transformed into:
\begin{small}
\begin{equation}
\begin{aligned}
    \min_{\delta} \quad & \mathop{\mathbb{E}}_{x \sim \mathcal{X}} [\mathcal{L}_{ASR}(f(\mathcal{D}(\mathcal{E}(x) + \delta)), t)] & \\
    \text{s.t.} \quad & \|\delta\|_\infty \le \tau, &
\end{aligned}
\end{equation}
\end{small}
where $\tau$ controls the $\ell_\infty$ boundary of the perturbation $\delta$.

Since the decoder satisfies the Lipschitz continuity~\cite{ye2019understanding,barrett2022certifiably}, we can ensure that the similarity of the output remains within a specified threshold with a particular probability when perturbations are added to the latent space. This indicates the following $r$-robustness probability bound.
\begin{theorem}\label{probability_bound}
Assume that the deterministic component of the decoder $\mathcal{D}$ is $a$-Lipschitz, and given two independent latent codes $z_1$ and $z_2$, then for $\forall r \in {\mathcal{R}^ + }$,
\begin{equation}
\mathcal{P}\left[ {{{\left\| {\mathcal{D}\left( {{z_1}} \right) - \mathcal{D}\left( {{z_2}} \right)} \right\|}_\infty } \le r} \right] \ge 1 - \min \left\{ {1,\frac{{{a^2}\tau }}{{{r^2}}}} \right\}.
\end{equation}
\end{theorem}

\begin{figure}[t]
    \centering
    \includegraphics[width=0.8\linewidth]{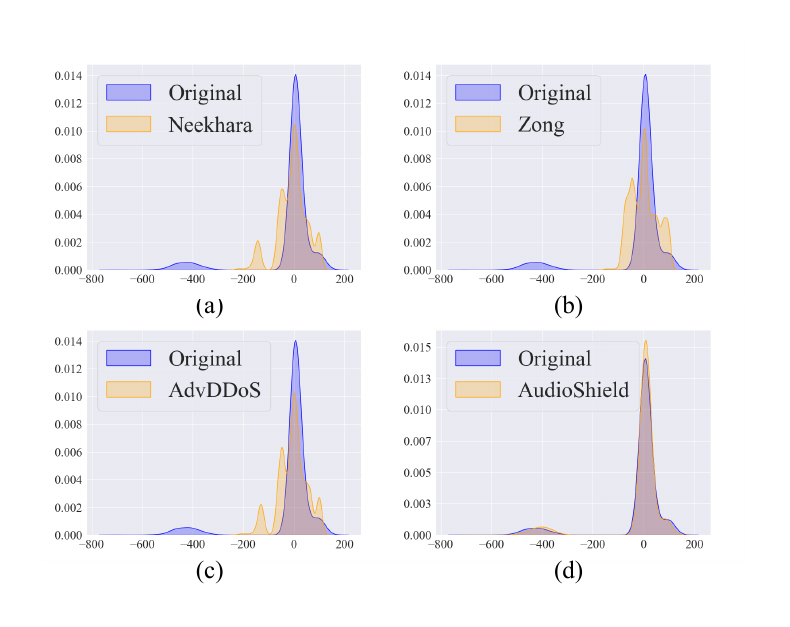}
    \caption{MFCC feature distributions of the original and adversarial audios generated by four methods. We use kernel density estimation (KDE) to calculate the distribution, with the horizontal axis representing MFCC mean values and the vertical axis representing density.}
    \label{fig:distribution}
    \vspace{-3mm}
\end{figure}

Considering that the autoencoder attempts to generate samples that are close to the original data distribution, Theorem~\ref{probability_bound} indicates that adding perturbations to the latent space will not significantly affect the distribution discrepancy between the output and input samples of the autoencoder, thus providing a guarantee for maintaining the audio quality. 
See Appendix~\ref{append:proof} for the proof. 
Besides, the MFCC feature distributions in Figure~\ref{fig:distribution} demonstrate that the adversarial examples generated by \UAPSystemName exhibit a distribution similar to that of the original audio. Conversely, the distributions of samples generated by other methods diverge significantly from the original audio distribution, thereby indicating a greater degree of noise.

Since robust features learned by UAPs should be input-agnostic, and the inputs have limited sample variability, we add Gaussian noise $p \sim \mathcal{N}(0, \sigma^2)$ with a variance of $\sigma^2$ to the UAP during training to increase the diversity of inputs and prevent overfitting or falling into local minima~\cite{kurakin2018adversarial,jiang2019black}. Therefore, the optimization problem becomes:
\begin{equation}
\begin{aligned}
    \min_{\delta} \quad & \mathop{\mathbb{E}}_{x \sim \mathcal{X}, p\sim \mathcal{N}(0,\sigma^2)} [\mathcal{L}_{ASR}(f(\mathcal{D}(\mathcal{E}(x) + \delta + p)), t)] & \\
    \text{s.t.} \quad & \|\delta\|_\infty \le \tau. &
\end{aligned}
\end{equation}

\subsection{Target Feature Adaptation}\label{subsec:target_feature_adaptation}

As mentioned earlier, the latent features of different target audios have different impacts on the transferability of UAPs. 
Given that the latent code contains rich acoustic and semantic features and that ASR models primarily focus on key features such as phonetic and semantic features, it is important to minimize the interference from other features, such as speaker identity. A straightforward approach might be to select a group of target audios with different styles and random speakers. Then, in each iteration, an audio is cherry picked from this group, fed into the encoder to extract latent features as the target code, and used for optimization. However, in practice, we find that this approach often leads to local optima, negatively affecting the optimization results. We speculate that the gradient descent direction is disrupted by multiple target audios, causing the gradient descent process to be less smooth, and ultimately reducing transferability. The above pilot study provides more justification for the target audio selection strategy in Section~\ref{subsec:protection_preparation}.

By minimizing interference from other features and irrelevant information and selecting the best target audio, the UAP can learn better target features in the optimization stage, and as a result, the generated adversarial audio is more likely to cross the decision boundary in the target model's decision space.
Specifically, we first generate a group of audio clips corresponding to the given target text $t$ using TTS, and choose the best target audio $x_t$ according to the aforementioned selection criteria. Then, we extract the latent features of the target audio using the encoder, denoted as $z_t = \mathcal{E}(x_t)$. Note that in the targeted setting, the selected target audio for all audios used to train UAPs is the same. 
Our optimization goal is to minimize the cosine similarity between the perturbation $\delta$ and $z_t$. Thus, the objective function can be expressed as:
\begin{equation}
\begin{aligned}
    \min_{\delta} \quad & \mathop{\mathbb{E}}_{x \sim \mathcal{X}, p\sim \mathcal{N}(0,\sigma^2)} [\mathcal{L}_{ASR}(f(\mathcal{D}(z_x + \delta + p)), t) & \\ &+ \lambda \cdot \mathcal{L}_{Sim}(z_t, \delta) ] & \\
    \text{s.t.} \quad & \|\delta\|_\infty \le \tau, &
\end{aligned}
\label{eq:final}
\end{equation}
where $\mathcal{L}_{Sim}$ represents the cosine similarity loss, $z_x = \mathcal{E}(x)$ represents the latent code of input audio $x$, $\lambda$ controls the contribution ratio between the two loss terms. 

Then, in each iteration, we train the UAPs in a mini-batch manner. To restrict the perturbation size, the perturbation is updated through Projected Gradient Decent (PGD)~\cite{madry2017towards}:
\begin{equation}
\delta \gets \text{clip}(\delta - \alpha \cdot \text{sign}(\nabla_{\delta} \mathcal{L}_{total}), -\tau, \tau),
\end{equation}
where $L_{total}$ denotes the above expectation of the total loss in a mini-batch, $\alpha $ denotes the learning rate, $\text{clip}$ denotes the clip operation to constrain the perturbations within $[-\tau, \tau]$.

\subsection{Physical Robust Perturbation}
In over-the-air scenarios, when audio adversarial examples are played through a speaker and transmitted through the air, they undergo inevitable and severe distortions, including signal attenuation, multipath effects, and environmental noise, which can diminish the effectiveness of the adversarial example. To address this, we integrate the effects of environmental absorption and reverberation into the perturbation optimization process by using room impulse responses (RIRs) to simulate the transfer function between the sound source and the microphone. This approach allows us to simulate air-induced distortions during the optimization of adversarial perturbations, thereby enhancing their robustness. 

Since RIRs vary significantly depending on the environment, we use a set of real RIRs collected in different environments for optimization. Specifically, we utilize the Aachen impulse response database~\cite{jeub2009binaural} and the MIT IR Survey~\cite{mit}, which together provide a total of 615 RIRs. By integrating these RIRs, Equation~\ref{eq:final} is transformed into:
\begin{equation}
\begin{aligned}
    \min_{\delta} \quad & \mathop{\mathbb{E}}_{x \sim \mathcal{X}, p\sim \mathcal{N}(0,\sigma^2), r\sim \mathcal{R}} [\mathcal{L}_{ASR}(f(\mathcal{D}(z_x + \delta + p)\otimes r), t) &\\ &+ \lambda  \cdot \mathcal{L}_{Sim}(z_t, \delta) ] & \\
    \text{s.t.} \quad & \|\delta\|_\infty \le \tau, &
\end{aligned}
\label{eq:final_OTA}
\end{equation}
where $\mathcal{R}$ represents the distribution of RIRs, $\otimes$ represents the convolution operation. The generated example $x'$, after being convolved with $r$, represents the example that has taken into account the simulated propagation effects.

\section{Experiments}
\subsection{Experiment Setup}\label{subsec:experiment_setup}
\noindent\textbf{Datasets.} We randomly select 500 and 2,000 audio-text pairs with durations between 1 to 5 seconds from the LibriSpeech~\cite{panayotov2015librispeech} and VCTK Corpus~\cite{yamagishi2019cstr}, respectively, as the training dataset and test dataset.

\noindent\textbf{Target Texts.} Similar to~\cite{chen2020devil,fang2024zero}, we choose 10 commonly used commands as the target texts: \textit{call my wife}, \textit{make it warmer}, \textit{navigate to my home}, \textit{open the door}, \textit{open the website}, \textit{play music}, \textit{send a text}, \textit{take a picture}, \textit{turn off the light}, and \textit{turn on airplane mode.}

\noindent\textbf{Target Models.} Consistent with competitors, we train \UAPName locally using DeepSpeech2~\cite{amodei2016deep} as the surrogate model. In the testing phase, to fully demonstrate the effectiveness of our method, we conduct experiments on 10 modern ASR systems. Specifically, for digital scenarios, we test on four commercial cloud ASR APIs (Google~\cite{google}, Amazon~\cite{amazon}, iFlytek~\cite{iflytek}, and Alibaba~\cite{alibaba}), two LLM-powered ASR (Qwen-Audio~\cite{chu2023qwenaudioadvancinguniversalaudio} and MooER~\cite{xu2024mooerllmbasedspeechrecognition}) and one state-of-the-art (SOTA) open-sourced NN-based ASR (OpenAI Whisper-large-v3~\cite{radford2023robust}). For physical robustness, we test on three voice assistants: Google Assistant~\cite{google_assistant}, Amazon Alexa~\cite{amazon_alexa}, and Apple Siri~\cite{apple_siri}.
In the over-the-line setting, we train UAPs on 2 NVIDIA GeForce RTX 4090 GPUs, running a 64-bit Ubuntu 18.04 operating system.

\noindent\textbf{Parameter Settings.} For the configuration, we set $\tau$ to 0.5, $\sigma$ to 1.0, $\lambda$ to 50, and a batch size to 16. We use Adam as the optimizer with a learning rate of 0.001. The impact of hyper-parameters will be analyzed in Section~\ref{sec:ablation}. 

\noindent\textbf{Competitors.} To demonstrate the superior performance of \UAPSystemName, we compare it with three SOTA methods: Neekhara \etal~\cite{neekhara2019universal}, Zong \etal~\cite{zong2021targeted}, and AdvDDoS~\cite{ge2023advddos}. All three methods generate UAPs for ASR tasks. The work done by Neekhara \etal~is tailored for untargeted scenarios, while the work done by Zong \etal~and AdvDDoS are only applicable to targeted scenarios. Additionally, save for universality, AdvDDoS also achieves transferability. Note that we train the UAPs for Neekhara \etal~ in an untargeted manner, so different target texts make no difference to the performance of Neekhara \etal~ in our experiments. We do not consider transfer-only competitors since they lack universality.

\begin{table}[t]
\centering
\caption{The recognition results on benign examples.}
\label{tab:benign}
\resizebox{0.6\linewidth}{!}{
\begin{tabular}{c|c|c|c|c}
\Xhline{1px}
\textbf{Metrics} & \textbf{Google} & \textbf{Amazon} & \textbf{iFlytek} & \textbf{Alibaba} \\ \Xhline{1px}
WER & 4.09 & 2.57 & 3.40 & 3.01 \\ \hline
CER & 3.78 & 3.09 & 3.29 & 3.03 \\ \Xhline{1px}
\end{tabular}
}
\vspace{-2mm}
\end{table}

\begin{table}[t]
\centering
\caption{Comparison 
on commercial ASR APIs.}
\label{tab:overall}
\resizebox{0.99\linewidth}{!}{
\begin{tabularx}{\textwidth}{c|>{\centering\arraybackslash}X>{\centering\arraybackslash}X>{\centering\arraybackslash}X|>{\centering\arraybackslash}X>{\centering\arraybackslash}X>{\centering\arraybackslash}X|>{\centering\arraybackslash}X>{\centering\arraybackslash}X>{\centering\arraybackslash}X|>{\centering\arraybackslash}X>{\centering\arraybackslash}X>{\centering\arraybackslash}X}
\Xhline{1px}
\multirow{2}{*}{\textbf{Method}} & \multicolumn{3}{c|}{\textbf{Google}} & \multicolumn{3}{c|}{\textbf{Amazon}} & \multicolumn{3}{c|}{\textbf{iFlytek}} & \multicolumn{3}{c}{\textbf{Alibaba}} \\ \cline{2-13}
 & \textbf{PSR} & \textbf{CER} & \textbf{WER} & \textbf{PSR} & \textbf{CER} & \textbf{WER}  & \textbf{PSR} & \textbf{CER} & \textbf{WER} & \textbf{PSR} & \textbf{CER} & \textbf{WER} \\ \Xhline{1px}
Neekhara \etal~\cite{neekhara2019universal} & 31.10 & 35.85 & 43.12 & 32.27 & 36.95 & 44.58 & 74.60 & \textbf{86.26} & \textbf{108.64} & 58.17 & \textbf{77.56} & 95.38 \\ \hline
Zong \etal~\cite{zong2021targeted} & 62.67 & 57.79 & 65.67 & 50.31 & 49.73 & 56.06 & 50.34 & 48.62 & 63.43 & 51.10 & 50.11 & 66.91 \\ \hline
AdvDDoS~\cite{ge2023advddos} & 31.02 & 36.21 & 43.82 & 19.19 & 26.17 & 30.80 & 39.85 & 42.19 & 55.28 & 63.76 & 75.78 & \textbf{101.11} \\ \hline
\UAPSystemName & \textbf{90.55} & \textbf{80.94} & \textbf{90.06} & \textbf{77.75} & \textbf{69.65} & \textbf{82.54} & \textbf{80.10} & 66.80 & 88.77 & \textbf{81.05} & 68.59 & 87.72 \\ \Xhline{1px}
\end{tabularx}
}
\vspace{-3mm}
\end{table}

\begin{table}[t]
\centering
\caption{Comparison of several transcription results from iFlytek API.}
\label{tab:examples}
\resizebox{0.95\linewidth}{!}{
\begin{tabular}{>{\centering\arraybackslash}m{3cm}|>{\centering\arraybackslash}m{3cm}|>{\centering\arraybackslash}m{3cm}|>{\centering\arraybackslash}m{3cm}|>{\centering\arraybackslash}m{3cm}}
\Xhline{1px}
\textbf{Original} & \textbf{Neekhara \etal}~\cite{neekhara2019universal} & \textbf{Zong \etal}~\cite{zong2021targeted} & \textbf{AdvDDoS}~\cite{ge2023advddos} & \textbf{\UAPSystemName} \\ \Xhline{1px}
I've not said anything to them, they know & I've not said anything to them they know & \textcolor{red}{has} not said anything to them they know & I've not said anything to them they know & \textcolor{red}{no I don't know who had} anything to \textcolor{red}{do with} \\ \hline
one season, they might do well & 1 season they might do well & they might \textcolor{red}{be} well & 1 \textcolor{red}{piece and you} might \textcolor{red}{be} well & \textcolor{red}{most of the time} \\ \hline
they have shown a great desire and attitude & \textcolor{red}{has} shown a great desire and attitude & \textcolor{red}{it's been} a great desire and attitude & a great desire and \textcolor{red}{that} & \textcolor{red}{all right so} \\ \hline
I decided it is going to be William & I decided it is going to be \textcolor{red}{later} & I decided \textcolor{red}{to return} to \textcolor{red}{england} & I decided it \textcolor{red}{was} going to be \textcolor{red}{there} & \textcolor{red}{no I have} it is going to be \textcolor{red}{all you} \\ \hline
Charles Kennedy had an effective outing & \textcolor{red}{still going to be hiding} & \textcolor{red}{that's going to be} & \textcolor{red}{that's going to be having} & \textcolor{red}{just kind of the} \\ \Xhline{1px}
\end{tabular}
}
\vspace{-4mm}
\end{table}

\noindent\textbf{Evaluation Metrics.} For the evaluation of protection effectiveness, we use the protection success rate (PSR, \%), character error rate (CER, \%), and word error rate (WER, \%) as metrics. Note that a protection is considered successful only when the example's CER reaches 50\% or higher. For these metrics, higher values indicate better performance. For audio quality, since we do not introduce noise to the original audio, the signal-to-noise ratio (SNR) adopted by traditional methods is not suitable for evaluating our method. Therefore, following~\cite{yu2023smack,yu2023antifake}, we use the SOTA DNN-based audio quality assessment system, NISQA~\cite{mittag2021nisqa}, which quantifies audio quality and naturalness on a scale from 1 to 5. A higher NISQA denotes higher audio quality.

\subsection{Evaluation on Cloud ASR APIs}
In an over-the-line setting, the generated audio adversarial examples are directly input into the target ASR APIs. First, we assess the functionality of the four commercial ASR APIs using benign audio clips. As shown in Table~\ref{tab:benign}, all ASR models can accurately recognize these audio clips, showing a low CER and a low WER, specifically lower than 5\%.

Then, the UAPs are trained in a targeted manner for three competitors with the target text being ``open the door'' before being added to all audio clips in the test stage (see Appendix~\ref{append:complete} for results on other target texts).
We observe that there are a few examples for which the results returned by the commercial ASR APIs are either empty or ``NA''. We consider these to be anomalous examples detected by the ASR systems, which is why no result is returned. Therefore, we filter out these examples.
The results of \UAPSystemName and competitors on commercial ASR APIs are presented in Table~\ref{tab:overall}. 
The mean PSR of \UAPSystemName reaches 82.36\%, surpassing Neekhara \etal~\cite{neekhara2019universal}, Zong \etal~\cite{zong2021targeted}, and AdvDDoS by 33.32\%, 28.75\%, and 43.90\%, respectively.
Concurrently, the adversarial examples generated by our method demonstrate superior transferability. Specifically, our method achieves high PSR across all 4 commercial ASR APIs, whereas the three competitors exhibit high PSR only on specific models but perform poorly on others. For instance, AdvDDoS reaches a 63.76\% PSR on Alibaba, while the PSR on the other three models remains below 40\%.

The results indicate that our method demonstrates superior transferability compared to other approaches, thereby showing stronger scalability to unknown ASR models.
It is worth noting that making a small perturbation effective across different ASR models' decision boundaries is quite challenging.
Moreover, Zong \etal~\cite{zong2021targeted} and AdvDDoS use a two-stage generation algorithm. In the first stage, no constraints are applied to the perturbation, generating a UAP that works on any audio. In the second stage, they gradually reduce the perturbation's magnitude to minimize noise while maintaining a success rate above a certain threshold. This approach is prone to getting stuck in local optima. In contrast, our method employs a single-stage optimization process, leveraging the latent features of the target audio to guide the gradient direction during iterations, thereby improving transferability. This explains why our method outperforms the competitors.
Furthermore, our method induces more transcription errors compared to competitors, as evidenced by the CER and WER of \UAPSystemName, which are 12.34\% and 14.34\% higher than those of the competitor with the highest error rates among the three, respectively. 

\begin{table}[t]
\centering
\caption{Comparison on two LLM-powered ASR and Whisper-large-v3.}
\label{tab:llm}
\resizebox{0.99\linewidth}{!}{
\begin{tabular}{c|c c c|c c c|c c c}
\Xhline{1px}
\multirow{2}{*}{\textbf{Method}} & \multicolumn{3}{c|}{\textbf{Qwen-Audio}} & \multicolumn{3}{c|}{\textbf{MooER}} & \multicolumn{3}{c}{\textbf{Whisper}} \\ \cline{2-10}
 & \textbf{PSR} & \textbf{CER} & \textbf{WER} & \textbf{PSR} & \textbf{CER} & \textbf{WER} & \textbf{PSR} & \textbf{CER} & \textbf{WER} \\ \Xhline{1px}
Neekhara \etal~\cite{neekhara2019universal} & 45.08 & 62.52 & 75.74 & 39.49 & 42.86 & 62.70 & 24.42 & 32.52 & 41.32 \\ \hline
Zong \etal~\cite{zong2021targeted} & 70.38 & \textbf{79.25} & 96.65 & 63.64 & 54.27 & 76.20 & 66.48 & 60.37 & 75.73 \\ \hline
AdvDDoS~\cite{ge2023advddos} & 45.00 & 60.85 & 76.88 & 47.93 & 45.57 & 67.08 & 36.21 & 39.61 & 50.56 \\ \hline
\UAPSystemName & \textbf{77.22} & 70.43 & \textbf{99.47} & \textbf{79.99} & \textbf{64.63} & \textbf{98.32} & \textbf{85.43} & \textbf{71.48} & \textbf{96.76} \\ \Xhline{1px}
\end{tabular}
}
\vspace{-3mm}
\end{table}

Table~\ref{tab:examples} presents several transcription results from several adversarial examples generated by each method. The sentences in the first column denote the original audio transcripts. The words that are different from the original transcripts in the transcription results generated by the four methods are highlighted in red. The results show that \UAPSystemName effectively causes most of the words in the original sentence to be transcribed incorrectly, significantly altering the original semantics. In contrast, competitors often fail to induce errors in certain words, which diminishes their practicality. For example, in the examples from the second row (excluding the header row), the transcription results of the three competitors all contain the original sentence's words ``might'' and ``well'', whereas the transcription caused by \UAPSystemName does not include any words from the original sentence. Overall, our method demonstrates strong competitiveness in fooling ASR APIs.
The waveforms and spectrograms of these examples can be found in Appendix~\ref{append:vision}.

\subsection{Evaluation on LLM-powered ASR}
To further evaluate the protection performance of our method, we conduct experiments on two LLM-powered ASR: Qwen-Audio~\cite{chu2023qwenaudioadvancinguniversalaudio} and MooER~\cite{xu2024mooerllmbasedspeechrecognition}. Qwen-Audio is a fundamental multi-task audio-language model that supports various tasks, languages, and audio types, serving as a universal audio understanding model. MooER is the SOTA LLM for audio understanding, which is trained on 80,000 hours of data. Additionally, we select Whisper-large-v3~\cite{radford2023robust}, a SOTA NN-based ASR system trained on a diverse audio dataset comprising 680,000 hours of audio, for evaluation. Although it is not LLM-powered, its recognition performance surpasses that of some LLM-powered ASR. Evaluating \UAPSystemName on these three models provides a comprehensive demonstration of its effectiveness on SOTA ASR.

In our experiments, we observe a few anomalous examples in the results returned by the ASR, where the output consists of a short string repeated a large number of times. We consider these to be recognition failures, similar to cases where commercial ASR models return empty or ``NA'' results; therefore, we filter them out. Notably, while all four methods exhibit some anomalous examples, \UAPSystemName has the fewest, reflecting its advantage in audio quality.
As shown in Table~\ref{tab:llm}, \UAPSystemName achieves the highest PSR and WER across three ASR models, with an average PSR of 80.88\% and an average WER of 98.18\%, demonstrating its superior protection performance. For competitors, the PSRs of Neekhara \etal \space and AdvDDoS do not exceed 50\% on any of the three models. The best-performing competitor is Zong \etal, with an average PSR of 66.83\%, but its audio quality is too poor, with an average NISQA score of only 1.14. In contrast, our method achieves higher NISQA score while still outperforming Zong \etal by 14.05\% in PSR, showcasing the advantage of our method in maintaining high audio quality while ensuring greater effectiveness.
It is also worth mentioning that \UAPSystemName shows the most stable performance when transferring from commercial ASR to LLM-powered ASR, with only a 1.48\% difference in average PSR between the two types, demonstrating strong transferability.

\subsection{Audio Quality Evaluation}\label{subsec:quality}
To comprehensively evaluate the advantages of adversarial examples generated by our method in terms of audio quality, we conduct both objective and subjective evaluations.

\subsubsection{Objective Evaluation}
As mentioned previously, we utilize NISQA~\cite{mittag2021nisqa} as the metric in the objective evaluation. In addition to assessing overall speech quality, NISQA predicts four specific dimensions of speech quality: noisiness, coloration, discontinuity, and loudness, offering a more comprehensive evaluation. 

\begin{table}[t]
\centering
\caption{Comparison of protection success rates on iFlytek API and objective audio quality.}
\label{tab:objective}
\resizebox{0.9\linewidth}{!}{
\begin{tabular}{c|c c|c c|c c}
\Xhline{1px}
\multirow{2}{*}{\textbf{Command}} & \multicolumn{2}{c|}{\textbf{Zong \etal}~\cite{zong2021targeted}} & \multicolumn{2}{c|}{\textbf{AdvDDoS}~\cite{ge2023advddos}} & \multicolumn{2}{c}{\textbf{\UAPSystemName}} \\ \cline{2-7}
 & \textbf{PSR} & \textbf{NISQA} & \textbf{PSR} & \textbf{NISQA} & \textbf{PSR} & \textbf{NISQA} \\ \Xhline{1px}
call my wife & 73.05 & 1.11 & 58.38 & 1.52 & 80.84 & 2.42 \\ \hline
make it warmer & 56.37 & 1.25 & 62.01 & 1.53 & 72.55 & 2.11 \\ \hline
navigate to my home & 47.42 & 1.00 & 64.61 & 1.63 & 68.26 & 2.31 \\ \hline
open the door & 50.34 & 1.33 & 39.85 & 1.54 & 80.10 & 2.45 \\ \hline
open the website & 28.14 & 1.03 & 55.75 & 1.53 & 69.73 & 2.09 \\ \hline
play music & 36.31 & 1.04 & 62.40 & 1.68 & 56.75 & 2.02 \\ \hline
send a text & 72.23 & 1.08 & 58.46 & 1.56 & 66.58 & 2.36 \\ \hline
take a picture & 40.95 & 1.34 & 81.20 & 1.37 & 55.86 & 2.19 \\ \hline
turn off the light & 64.08 & 1.04 & 57.43 & 1.32 & 73.45 & 2.55 \\ \hline
turn on airplane mode & 62.56 & 1.22 & 67.44 & 1.15 & 81.85 & 2.33 \\ \hline
\textbf{Average} & \textbf{53.15} & \textbf{1.14} & \textbf{60.75} & \textbf{1.48} & \textbf{70.60} & \textbf{2.28} \\ \Xhline{1px}
\end{tabular}
}
\vspace{-3mm}
\end{table}

Table~\ref{tab:objective} reports the PSR and NISQA results for different commands. We do not provide results for Neekhara \etal~\cite{neekhara2019universal} since their work is in an untargeted manner. Results show that \UAPSystemName achieves the highest NISQA under each target text, although it still does not match the NISQA of benign audio (3.99$\pm$0.61). Nevertheless, the average NISQA of our method is twice that of Zong \etal \space and 0.8 higher than that of AdvDDoS. More importantly, our method not only achieves high audio quality, but also maintains a high PSR, with an average PSR of 70.60\%, which is 17.45\% and 9.85\% higher than that of Zong \etal \space and AdvDDoS. These results fully demonstrate the superiority of our method, as it enhances both audio quality and protection performance.

\subsubsection{Subjective Evaluation}
We conduct a user study on the subjection evaluation of our proposed method. To be concrete, we published a survey on Amazon Mechanical Turk~\cite{amazon_turk}, a crowdsouring platform, to subjectively evaluate the audio quality of adversarial audios generated by all methods. This study was approved by the institutional review board (IRB), and we followed best practice for ethical human subjects survey research. 
We recruited 53 participants from the USA and Australia, aged 18 to 30, who have normal hearing and demonstrate adequate proficiency in English. All participants agreed that their responses can be used for academic research. We removed four junk responses that gave the same score to all audios, and finally obtained 49 valid responses. Additionally, we did not collect any personal information related to the participants. In our study, we carefully selected audio clips corresponding to neutral and commonly used ground-truth texts to minimize bias and discrimination. Each participant was paid \$1.0 for each question, except for junk responses.

\noindent\textbf{Survey Protocol.}
We selected five audio clips each from benign audio and the audio generated by our method and three competitors (with the target text being ``open the door''), totaling 25 clips. These clips were shuffled in advance to avoid bias. Note that, before the study, the participants had no knowledge of whether these clips contain clean audios or adversarial audios. In each question, an audio clip was played and participants were asked to rate each clip using the Mean Opinion Score (MOS), based on a Likert scale~\cite{likert1932technique} ranging from 1 to 5, where 1 indicates very poor quality and 5 indicates very good quality, with intermediate values representing varying levels of quality. The survey questions are provided in Appendix~\ref{append:user_study}.

\noindent\textbf{Survey results.}
As a reference, we also provide the NISQA for all 2,000 audio clips. The NISQA scores for Neekhara \etal, Zong \etal, AdvDDoS and \UAPSystemName are 1.71, 1.33, 1.54 and 2.45, and the MOSs in the subjective evaluation for these four methods are 1.56, 1.16, 1.44, 2.91.
Similar to the objective evaluation, the adversarial examples generated by our method also receive a higher MOS in the subjective evaluation, surpassing the three competitors by 1.35, 1.75, and 1.47, respectively. 
Additionally, we conduct a statistical analysis using the Mann-Whitney U-test~\cite{mann1947on}, with the null hypothesis asserting that AudioShield's MOS is not significantly higher than that of the competitors'. The null hypothesis for the three competitors is rejected at $p$-values of $2.68 \times 10^{-55}$, $3.34 \times 10^{-78}$, and $8.40 \times 10^{-60}$, with a significance level of 0.05.
The results strongly indicate that the audio quality produced by our method is significantly superior to that produced by competitors and supports our earlier analysis. Specifically, by adding perturbations in the latent space rather than directly introducing noise in the acoustic space, our method effectively enhances audio quality.

\subsection{Evaluation on Long Audios}
To thoroughly evaluate the effectiveness of \UAPSystemName when the input of ASR is long audios, we conduct experiments using a long audio set, which consists of 400 audio clips ranging from 8 to 10 seconds. They are randomly selected from the VCTK Corpus~\cite{yamagishi2019cstr}, with an average of 21 words per clip. For our method, we tile the \UAPName to match the length of the latent code of the test audio. For competitors, we tile the UAPs to match the length of the test audio. We still use ``open the door'' as the target text. 

\begin{table}[t]
\centering
\caption{Comparison of protection performance on long audios.}
\label{tab:long}
\resizebox{0.9\linewidth}{!}{
\begin{tabular}{c|c c|c c|c c|c c|c}
\Xhline{1px}
\multirow{2}{*}{\textbf{Method}} & \multicolumn{2}{c|}{\textbf{Google}} & \multicolumn{2}{c|}{\textbf{Amazon}} & \multicolumn{2}{c|}{\textbf{iFlytek}} & \multicolumn{2}{c|}{\textbf{Alibaba}} & \multirow{2}{*}{\textbf{NISQA}} \\ \cline{2-9}
 & \textbf{PSR} & \textbf{WER} & \textbf{PSR} & \textbf{WER} & \textbf{PSR} & \textbf{WER} & \textbf{PSR} & \textbf{WER} \\ \Xhline{1px}
Neekhara \etal~\cite{neekhara2019universal}  & 22.00 & 38.12 & 25.00 & 48.65 & 69.27 & \textbf{89.86} & 35.43 & 53.15 & 2.28 \\ \hline
Zong \etal~\cite{zong2021targeted} & 63.59 & 71.62 & 49.61 & 59.34 & 53.55 & 67.29 & 51.15 & 67.04 & 1.25 \\ \hline
AdvDDoS~\cite{ge2023advddos} & 30.36 & 48.23 & 22.50 & 34.84 & 49.37 & 62.18 & 58.40 & \textbf{101.22} & 1.32 \\ \hline
\UAPSystemName & \textbf{82.14} & \textbf{83.68} & \textbf{76.69} & \textbf{76.95} & \textbf{82.07} & 84.82 & \textbf{90.40} & 89.67 & \textbf{2.43} \\ \Xhline{1px}
\end{tabular}
}
\end{table}

\begin{table}[t]
\centering
\caption{Comparison of protection success rates on voice assistants at different distances.}
\label{tab:ota}
\resizebox{0.9\linewidth}{!}{
\begin{tabular}{c|c c c|c c c|c c c}
\Xhline{1px}
\multirow{2}{*}{\textbf{Method}} & \multicolumn{3}{c|}{\textbf{Google Assistant}} & \multicolumn{3}{c|}{\textbf{Amazon Alexa}} & \multicolumn{3}{c}{\textbf{Apple Siri}} \\ \cline{2-10}
 & \textbf{10cm} & \textbf{20cm} & \textbf{50cm} & \textbf{10cm} & \textbf{20cm} & \textbf{50cm} & \textbf{10cm} & \textbf{20cm} & \textbf{50cm} \\ \Xhline{1px}
Neekhara \etal~\cite{neekhara2019universal} & 3/10 & 5/10 & 5/10 & 4/10 & 3/10 & 0/10 & 5/10 & 4/10 & 3/10 \\ \hline
Zong \etal~\cite{zong2021targeted} & 2/10 & 2/10 & 2/10 & 0/10 & 0/10 & 0/10 & 4/10 & 3/10 & 3/10 \\ \hline
AdvDDoS~\cite{ge2023advddos} & 7/10 & 6/10 & 5/10 & 4/10 & 5/10 & 2/10 & 6/10 & 4/10 & 2/10 \\ \hline
\UAPSystemName & 10/10 & 7/10 & 5/10 & 4/10 & 5/10 & 3/10 & 10/10 & 10/10 & 8/10 \\ \Xhline{1px}
\end{tabular}
}
\vspace{-3mm}
\end{table}

Table~\ref{tab:long} presents the protection performance on long audio clips. Our method achieves the highest PSR across all four commercial ASR APIs, specifically 82.14\%, 76.69\%, 82.07\%, and 90.40\%, with consistently high WER as well. Notably, compared to competitors, our method not only exhibits the best protection performance but also achieves the highest audio quality, with NISQA scores surpassing the three competitors by 0.15, 1.18, and 1.11, respectively. Furthermore, our method maintains strong protection performance on long audios, comparable to the results observed on short audios previously. In contrast, competitors exhibit decreased protection effectiveness on longer audio clips, highlighting the superior generalizability of \UAPSystemName. For example, the PSR for Neekhara \etal \space reduces from 58.17 to 35.43 when testing on Alibaba. These findings underscore the comprehensive advantage of \UAPName in fooling ASR models and suggest that protectors can reduce training costs by generating short \UAPName for application to longer audios.

\subsection{Over-the-Air Robustness Evaluation}
\begin{figure}[t]
  \centering
  \subfigure{
    \includegraphics[width=0.21\textwidth]{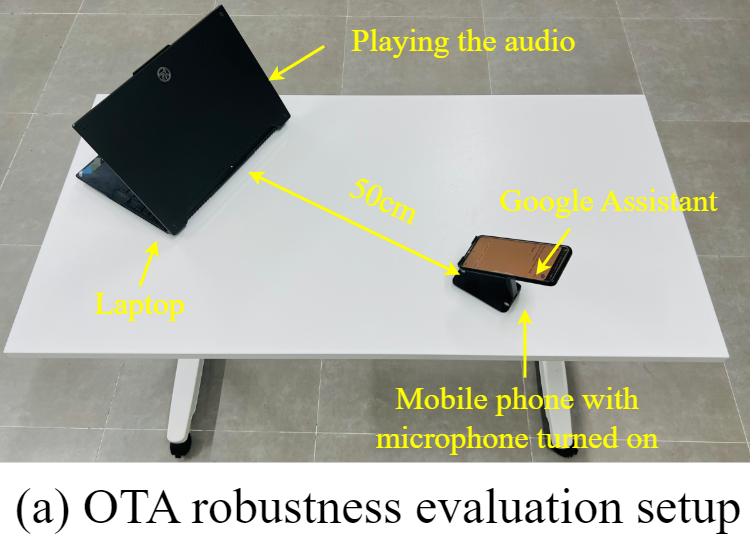}
   \label{fig:ota}
  }
  \subfigure{
    \includegraphics[width=0.23\textwidth]{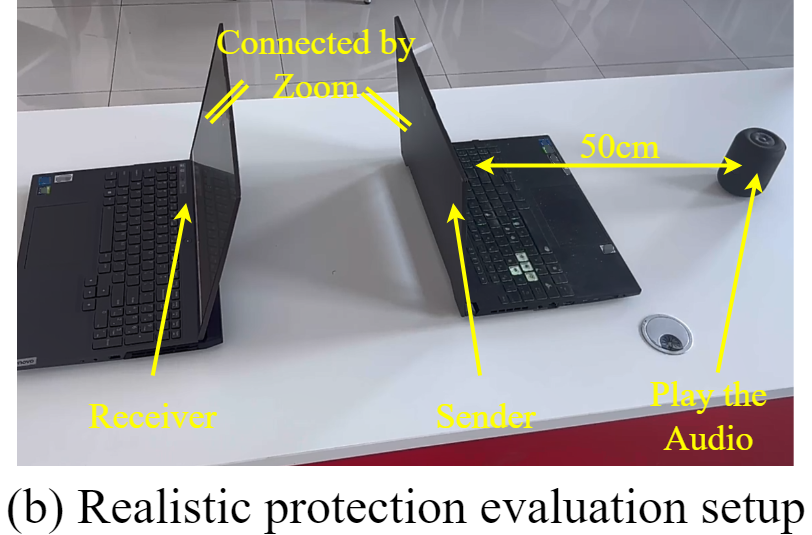}
    \label{fig:realistic}
  }
  \vspace{-1mm}
  \caption{Illustration of physical experimental setups.}
  \vspace{-5mm}
\end{figure}

To validate the robustness of the adversarial examples generated by \UAPSystemName in physical environments, we conduct experiments on three commonly used voice assistants: Google Assistant~\cite{google_assistant}, Amazon Alexa~\cite{amazon_alexa}, and Apple Siri~\cite{apple_siri}, in the over-the-air scenario. 
The experimental setup is shown in Figure~\ref{fig:ota}.
We use the speakers of an ASUS TUF Dash F15 laptop to play the audio adversarial examples, while the microphones of a Redmi K30S, with Google Assistant and Amazon Alexa apps installed, and an iPhone 13 Pro with Apple Siri are used to capture the audio.

For each method, we select 10 adversarial examples, which are generated by the corresponding original audio clips. Following previous work~\cite{ge2023advddos}, each audio is played three times during the test, and we record the best protection results. An example is considered successfully protected only if the assistant provides an output (in some cases, there is no output due to the large noise in audios generated by competitors) and the CER between the ground-truth text and the output transcription exceeds 50\%. The experiments take place in a room measuring $5.44m \times 3.56m \times 2.99m$. To assess the robustness of the protection regarding transmission distance, we conduct experiments at different distances between the speaker and the voice assistant.

As shown in Table~\ref{tab:ota}, the experimental results demonstrate the superiority of our method. On Google Assistant and Apple Siri, our average PSR reaches 73.33\% and 93.33\%, respectively, which are 13.33\% and 53.33\% higher than the best-performing competitor, AdvDDoS. Overall, \UAPSystemName outperforms competitors in protection performance in the physical domain. Specifically, on Google Assistant, at shorter distances (10cm and 20cm), our PSRs surpass those of competitors, achieving 100\% (10/10) at 10cm. At a longer distance (50cm), \UAPSystemName matches the best-performing competitor with a 50\% PSR. On Apple Siri, \UAPSystemName far exceeds the competitor at all distances, achieving a 100\% (10/10) PSR at 10cm and 20cm, and still maintaining an 80\% PSR at 50cm, while competitors only reach a maximum of 30\% PSR at 50cm. On Amazon Alexa, our long-distance PSR surpasses all competitors, while at short distances, \UAPSystemName matches the best-performing competitor (4/10 at 10cm and 5/10 at 20cm).

These results align with expectations. As the propagation distance of the examples in the air increases, the interference from environmental white noise also increases, leading to a decrease in PSRs. However, an interesting phenomenon worth noting is that in some cases, such as Neekhara \etal \space on Google and AdvDDoS on Alexa, the PSR at 10cm is actually lower than at 20cm. Upon closer observation, we find that in certain examples, the voice assistant successfully recognizes the input at 10cm, but correctly transcribes most of the words, leading to a failed protection. However, when the distance increases to 20cm, the influence of environmental white noise causes fewer words to be correctly transcribed after successful recognition, resulting in a higher CER and thus a successful protection. In contrast, our method performs well in all voice assistants, indicating that our method is more resilient to environmental noise compared to competitors.

Additionally, we observe that on Alexa, the total number of successful protected examples for all methods is the lowest among the three target voice assistants. We believe that this is related to the recognition capability of the voice assistant itself. Notably, Zong \etal \space achieves a 0\% PSR at all distances on Alexa - 
we find that Alexa does not produce any output for these examples. This is primarily due to the excessive noise in these examples, which makes it impossible for Alexa to recognize the human voice within them. In fact, most of the cases where no output is given for adversarial examples generated by competitors are due to excessive noise and poor quality, resulting in the voice assistant not being able to recognize the input. This further highlights the superiority of our method in terms of audio quality.

\subsection{Realistic Protection Evaluation}
\begin{table}[t]
\centering

\caption{Comparison of end-to-end protection performance in real-time scenarios.}
\label{tab:e2e}
\resizebox{0.95\linewidth}{!}{
\begin{tabular}{c|c c|c c|c c|c c|c|c|c}
\Xhline{1px}
\multirow{2}*{\textbf{Method}} & \multicolumn{2}{c|}{\textbf{Google}} & \multicolumn{2}{c|}{\textbf{Amazon}} & \multicolumn{2}{c|}{\textbf{iFlytek}} & \multicolumn{2}{c|}{\textbf{Alibaba}} & \multirow{2}{*}{\textbf{NISQA}} & \multirow{2}{*}{\textbf{MOS}} & \multirow{2}{*}{\textbf{Latency~(ms)}} \\ \cline{2-9}
 & \textbf{PSR} & \textbf{CER} & \textbf{PSR} & \textbf{CER} & \textbf{PSR} & \textbf{CER} & \textbf{PSR} & \textbf{CER} & & \\ 
\Xhline{1px}
Neekhara \etal~\cite{neekhara2019universal} & 1/10 & 20.05 & 1/10 & 12.22 & 7/10 & 57.76 & 1/10 & 19.55 & 1.37 & 1.95 & \textbf{5.91} \\ \hline
Zong \etal~\cite{zong2021targeted}     & 4/10 & 79.77 & 8/10 & \textbf{71.01} & 8/10 & 65.67 & 8/10 & 57.71 & 1.16 & 1.28 & 6.58 \\ \hline
AdvDDoS~\cite{ge2023advddos}         & 2/10 & 22.22 & 1/10 & 14.75 & 6/10 & 55.81 & 2/10 & 30.89 & 1.07 & 1.80 & 6.10 \\ \hline
\UAPSystemName     & \textbf{8/10} & \textbf{82.27} & \textbf{9/10} & 68.22 & \textbf{9/10} & \textbf{83.32} & \textbf{9/10} & \textbf{73.74} & \textbf{3.54} & \textbf{3.12} & 409.14 \\ 
\Xhline{1px}
\end{tabular}
}
\vspace{-3mm}
\end{table}

\noindent\textbf{End-to-End Evaluation.} To demonstrate the performance of \UAPSystemName in real-world scenarios, we conduct a real-time experiment. The experimental setup is shown in Figure~\ref{fig:realistic}. A Newmine BT51 Bluetooth speaker continuously plays audio while the microphone of ASUS TUF Dash F15 laptop receives the audio in real-time at a distance of 50cm. The audio is then processed by \UAPSystemName on the device, output to a virtual microphone, and the downstream communication software selects the virtual microphone as the input device. We select 10 audio clips for the real-time experiment and conduct an end-to-end evaluation of the entire process, using commercial APIs for recognition. We perform both objective and subjective evaluations of the audio. The experimental results are shown in Table~\ref{tab:e2e}.
Since our method introduces an additional autoencoder inference process, while competitors only need to add perturbations directly to raw audio, this leads to a significantly higher latency for \UAPSystemName. However, we believe that the 409.14ms delay is still within an acceptable range and meets real-time requirements.

Although our approach exhibits certain limitations in terms of latency, the audio quality of competitors is generally poor, with the highest NISQA and MOS only reaching 1.37 and 1.95, respectively, causing a loss of normal usability. In contrast, our method maintains relatively high audio quality, with an NISQA score of 3.54 and a MOS of 3.12. In terms of protection performance, \UAPSystemName significantly outperforms all competitors, achieving an average PSR of 87.50\% and an average CER of 76.89\%. Among competitors, only Zong \etal \space show a relatively good protection effect, reaching an average PSR of 70.00\%, while Neekhara \etal \space and AdvDDoS achieve average PSRs of only 25.00\% and 27.50\%, respectively. Considering overall protection performance, audio quality, and latency, we believe that our method still holds a significant advantage, especially in terms of high audio quality and strong protection performance.

\noindent\textbf{Impact of Various Environmental Factors.} We evaluate the impact of various factors on \UAPSystemName, testing it at the same 50cm distance in three different environments (bedroom, meeting room, outdoor), on different devices (laptop, mobile phone), and with different speakers (Speakers~A,B, and~C). We use Alibaba API and Qwen-Audio for recognition, and the results are presented in Table~\ref{tab:realistic}. Different environments primarily affect the propagation process of acoustic signals. Different receiving devices influence audio attributes, and different speakers also affect the volume, clarity, and other aspects of the audio. As a result, \UAPSystemName exhibits some fluctuations. For example, in the bedroom, the average PSR of Speaker~C in Qwen-Audio is the lowest, at only 25.00\%, whereas outdoors, it increases to 75.00\%, the highest for Speaker~C. We consider these fluctuations to be within a reasonable range, caused by the randomness introduced by various environmental factors. Overall, \UAPSystemName shows strong protective effects across different environments, with average protection success rates of 73.33\%, 93.33\%, and 76.67\% on Alibaba. Additionally, for different receiving devices, the average PSR for the laptop and mobile phone is 74.44\% and 77.22\%, respectively. For different speakers, the average PSR values are 84.17\%, 75.00\%, and 68.33\%, indicating that \UAPSystemName offers strong protection for both different receiving devices and speakers. In summary, these results demonstrate the robustness and adaptability of \UAPSystemName in physical protection scenarios.

\begin{table}[t]
\centering
\caption{Comparison of protection performance under different realistic settings.}
\label{tab:realistic}
\resizebox{0.95\linewidth}{!}{
\begin{tabular}{c|c|c c|c c|c c|c c|c}
\Xhline{1px}
\multirow{3}{*}{\centering \textbf{Environment}} & \multirow{3}{*}{\centering \textbf{Speaker}} & \multicolumn{4}{c|}{\textbf{Alibaba}} & \multicolumn{4}{c|}{\textbf{Qwen-Audio}} & \multirow{3}{*}{\centering \textbf{dB}} \\ \cline{3-10}
 &  & \multicolumn{2}{c|}{\textbf{laptop}} & \multicolumn{2}{c|}{\textbf{mobile phone}} & \multicolumn{2}{c|}{\textbf{laptop}} & \multicolumn{2}{c|}{\textbf{mobile phone}} & \\ \cline{3-10}
 &  & \textbf{PSR} & \textbf{CER} & \textbf{PSR} & \textbf{CER} & \textbf{PSR} & \textbf{CER} & \textbf{PSR} & \textbf{CER} & \\ 
\Xhline{1px}
\multirow{3}{*}{\centering bedroom} & Speaker~A & 10/10 & 84.54 & 8/10 & 55.63 & 10/10 & 93.87 & 5/10 & 46.61 & \multirow{3}{*}{\centering 40.2} \\ \cline{2-10}
 & Speaker~B & 6/10 & 58.86 & 9/10 & 61.59 & 4/10 & 50.30 & 7/10 & 63.55 & \\ \cline{2-10}
 & Speaker~C & 6/10 & 45.65 & 3/10 & 41.25 & 2/10 & 45.94 & 3/10 & 37.89 & \\ 
\Xhline{1px}
\multirow{3}{*}{\centering meeting room} & Speaker~A & 10/10 & 81.69 & 9/10 & 70.03 & 10/10 & 84.81 & 9/10 & 62.33 & \multirow{3}{*}{\centering 38.4} \\ \cline{2-10}
 & Speaker~B & 9/10 & 68.48 & 10/10 & 78.27 & 6/10 & 57.94 & 9/10 & 74.80 & \\ \cline{2-10}
 & Speaker~C & 9/10 & 59.12 & 10/10 & 80.99 & 9/10 & 69.79 & 9/10 & 71.36 & \\ 
\Xhline{1px}
\multirow{3}{*}{\centering outdoor} & Speaker~A & 8/10 & 57.06 & 8/10 & 64.67 & 9/10 & 55.29 & 5/10 & 57.51 & \multirow{3}{*}{\centering 51.3} \\ \cline{2-10}
 & Speaker~B & 9/10 & 65.69 & 8/10 & 70.67 & 4/10 & 47.34 & 9/10 & 65.80 & \\ \cline{2-10}
 & Speaker~C & 6/10 & 49.63 & 10/10 & 67.08 & 7/10 & 53.45 & 8/10 & 67.80 & \\ 
\Xhline{1px}
\end{tabular}
}
\vspace{-3mm}
\end{table}

\noindent\textbf{Case Study.} We conduct a case study using Zoom as the downstream communication software. Two devices (one for sending and one for receiving) connect remotely via Zoom, as shown in Figure~\ref{fig:realistic}. The sender's setup is consistent with that in the ``End-to-End Evaluation'', and the receiver (a Lenovo Y9000P laptop) records the remotely received audio for testing. The average PSR and CER reach 80.00\% and 66.74\%, respectively, demonstrating outstanding protection performance. Additionally, we recruited 36 participants to subjectively evaluate the audio quality, with an average MOS of 3.61, which is even higher than the  MOS (3.12) in the ``End-to-End Evaluation'', indicating high audio quality. In conclusion, our experiment strongly demonstrates the effectiveness of AudioShield in real-world deployment.

\subsection{Ablation Study}\label{sec:ablation}
We also explore several key components of training \UAPName and their impact on protection performance and audio quality. 

\noindent\textbf{Contribution of Target Feature Adaptation.}
Recall that we employ target feature adaptation to enable \UAPName to learn the latent features of the target text, which benefits the transferability of \UAPName. To verify the effectiveness of this component, we select ``open the door'' as the target text and test the protection performance of \UAPName trained using \one~only ASR Loss $\mathcal{L}_{ASR}$ and \two~both ASR Loss $\mathcal{L}_{ASR}$ and cosine similarity loss of latent features $\mathcal{L}_{Sim}$. 

\begin{figure}[t]
\centering
\includegraphics[width=0.8\linewidth]{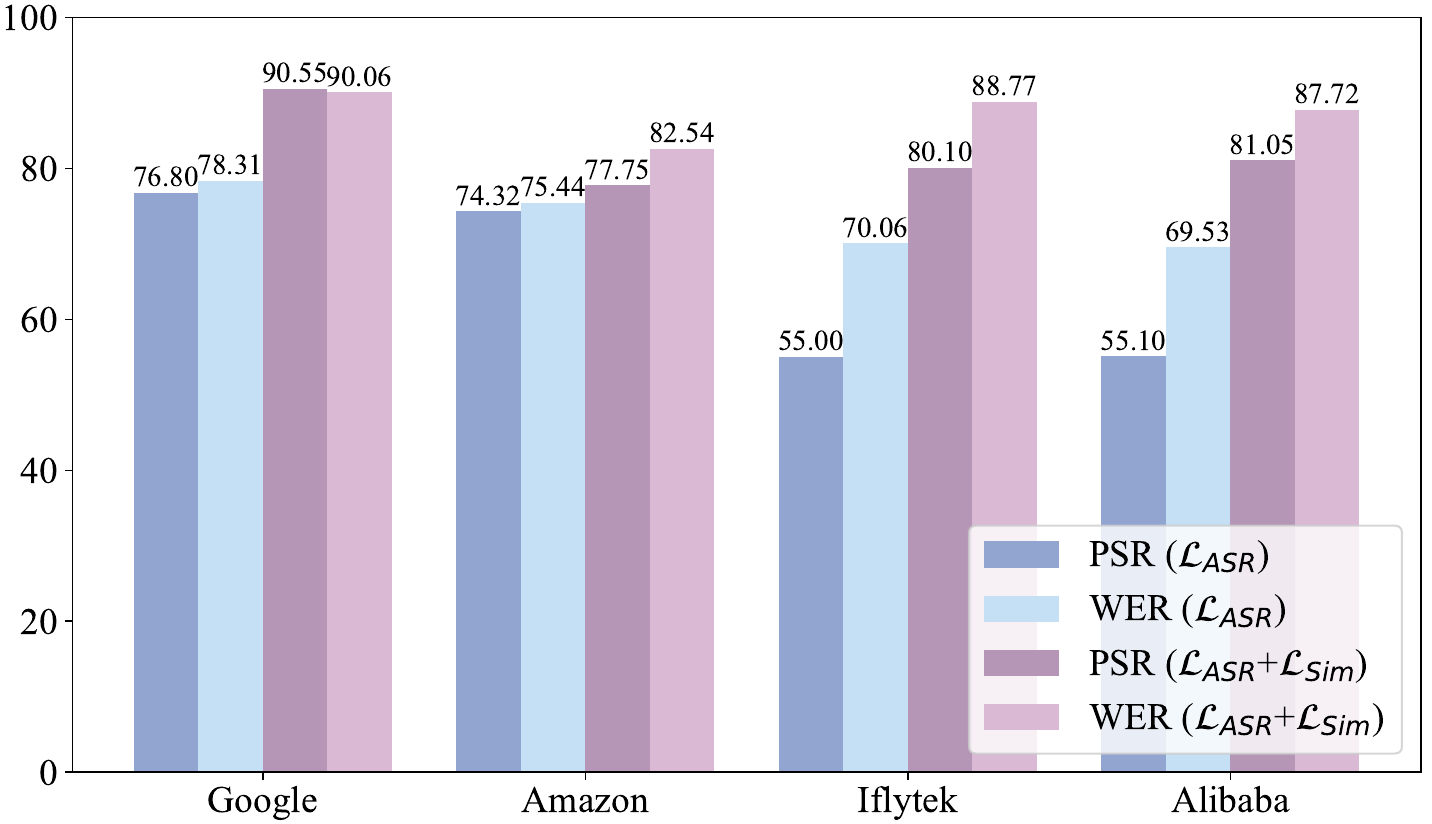}
\caption{Performance with/without target feature adaptation.}
\label{fig:ablation_loss}
\vspace{-5mm}
\end{figure}

According to Figure~\ref{fig:ablation_loss},
\UAPSystemName already achieves a high WER before implementing target feature adaptation, causing most words in the sentence to be incorrectly transcribed. However, its PSR on iFlytek and Alibaba remains around 55\%. After incorporating $\mathcal{L}_{Sim}$, the PSR of \UAPSystemName improves across all four commercial models, surpassing 75\%, with WER also showing an increase. This confirms our hypothesis that enabling \UAPName to learn the latent features of the target text is effective in enhancing its transferability.

\begin{figure}[t]
  \centering
  \subfigure{
    \includegraphics[width=0.25\textwidth]{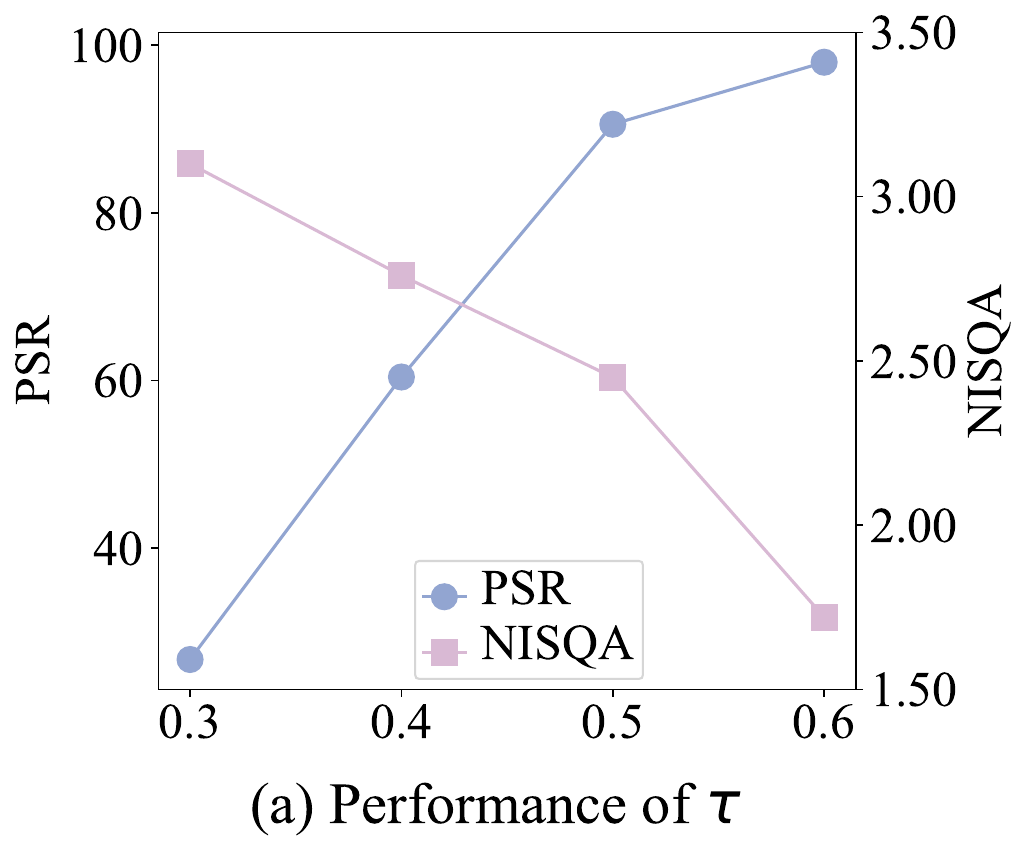}
    \label{fig:tau}
  }
  \subfigure{
    \includegraphics[width=0.20\textwidth]{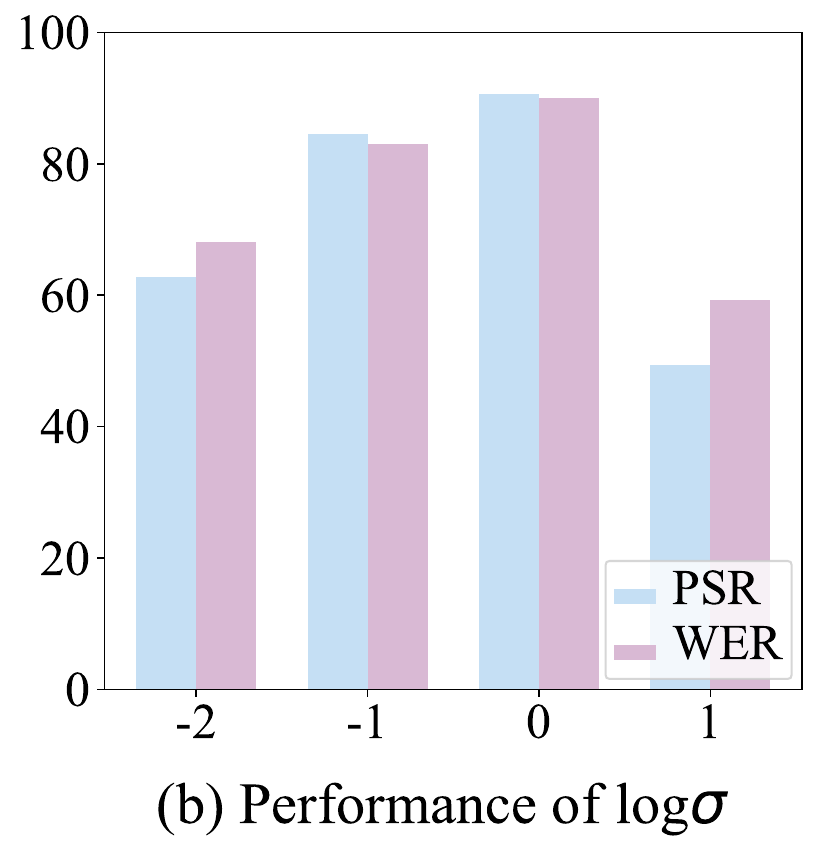}
    \label{fig:sigma}
  }
  \vspace{-5mm}
  \caption{Impact of hyper-parameters $\tau$ and $\sigma$.}
  \label{fig:hyparams}
  \vspace{-1mm}
\end{figure}

\noindent\textbf{Analyses of Hyper-parameters.}
Our method involves two important parameters, $\tau$ and $\sigma$. $\tau$ controls the perturbation boundary, which is crucial for balancing the trade-off between protection performance and audio quality. $\sigma$ controls the amount of Gaussian noise added during the training of \UAPName. To analyze the impact of these hyper-parameters, we set $\tau$ and $\sigma$ to different values and evaluate the performance on the Google API while keeping other variables constant. The experimental results are shown in Figure~\ref{fig:hyparams}.

As shown in Figure~\ref{fig:tau}, within the range of 0.3 to 0.6, PSR increases as $\tau$ increases, while NISQA gradually decreases. This indicates that a larger perturbation boundary results in higher protection performance but lower audio quality, whereas a smaller boundary leads to the opposite outcome. When $\tau = 0.5$, the PSR exceeds 90\%, and NISQA remains at a relatively high level of 2.45. However, when $\tau$ rises to 0.6, NISQA drops below 2, reaching only 1.72. Since our goal is to achieve protection effectiveness while maintaining high audio quality, we select $\tau = 0.5$ in our experiments. Figure~\ref{fig:sigma} presents the results of PSR and WER for different values of $\sigma$, with $\sigma$ set at four different magnitudes: 0.01, 0.1, 1.0, and 10. The overall curve shows an initial increase followed by a decline, peaking at $\sigma = 1.0$ ($\log \sigma = 0$), where PSR and WER reach 90.55\% and 90.06\%, respectively. Based on these results, we select $\sigma = 1.0$ in our experiments.

\subsection{Discussion on Countermeasures}
% \subsection{Countermeasures}
\label{subsec:countermeasures}
Once the eavesdroppers realize that the transcriptions they obtain are not what the speakers convey, they may seek to countermeasures to resist the protection for user's speech. Therefore, we explore the resilience of \UAPSystemName against three common and three adaptive countermeasures.

\begin{table}[t]
\centering
\caption{Results of \UAPSystemName against countermeasures.}
\label{tab:against_countermeasures}
\resizebox{0.95\linewidth}{!}{
\begin{tabular}{c|c|c c|c c|c c|c c|c}
\Xhline{1px}
\multirow{2}{*}{\textbf{Countermeasure}} & \multirow{2}{*}{\textbf{Setting}} & \multicolumn{2}{c|}{\textbf{Google}} & \multicolumn{2}{c|}{\textbf{Amazon}} & \multicolumn{2}{c|}{\textbf{iFlytek}} & \multicolumn{2}{c|}{\textbf{Alibaba}} & \multirow{2}{*}{\textbf{NISQA}} \\ \cline{3-10}
 &  & \textbf{PSR} & \textbf{WER} & \textbf{PSR} & \textbf{WER} & \textbf{PSR} & \textbf{WER} & \textbf{PSR} & \textbf{WER} \\ \Xhline{1px}
\multirow{3}{*}{\textbf{Local Smoothing}} & $h$ = 1 & 89.94 & 88.84 & 91.18 & 89.43 & 74.52 & 84.41 & 81.81 & 87.85 & 2.69 \\ \cline{2-11}
 & $h$ = 2 & 90.34 & 88.71 & 89.00 & 87.43 & 74.90 & 85.07 & 80.40 & 87.84 & 2.55 \\ \cline{2-11}
 & $h$ = 3 & 90.95 & 90.06 & 89.53 & 88.16 & 74.81 & 85.28 & 80.46 & 88.41 & 2.34 \\ \hline
\multirow{3}{*}{\textbf{Downsampling}} & DR = 14kHz & 88.64 & 87.69 & 89.88 & 87.77 & 75.35 & 84.58 & 77.42 & 86.70 & 2.24 \\ \cline{2-11}
 & DR = 12kHz & 89.54 & 88.32 & 90.75 & 88.70 & 76.34 & 87.52 & 82.86 & 88.72 & 2.33 \\ \cline{2-11}
 & DR = 10kHz & 89.32 & 89.11 & 91.21 & 89.06 & 81.71 & 92.31 & 86.10 & 90.51 & 2.57 \\ \Xhline{1px}
\end{tabular}
}
\vspace{-3mm}
\end{table}

\noindent\textbf{Local Smoothing.}
Local smoothing applies a sliding window that can easily eliminate small perturbations in carefully crafted adversarial examples. 
Given a sliding window, it calculates the average of sample points within a range of $2h$ before and after the audio sample point and replaces the current sample point with this average value. 
Specifically, if the current sample point is $x_i$, its value is replaced with the average of $\{x_{i-h}, \ldots, x_i, \ldots, x_{i+h}\}$. 
We evaluate the robustness of our method against local smoothing by setting $h$ to 1, 2, and 3. 
Results in Table~\ref{tab:against_countermeasures} indicate that \UAPSystemName exhibits strong resistance to local smoothing, as PSR remains consistently high in different settings without significant variation, while high audio quality is also maintained.

\noindent\textbf{Downsampling.}
Downsampling resists adversarial examples by removing the high-frequency components of the audio, as adversarial examples often add perturbations in these regions. Specifically, low-pass filtering and decimation are first applied to remove any high-frequency components above the Nyquist frequency, and the sampling rate is reduced. The audio is then upsampled back to the original sampling rate. Given that the original sampling rate of the audio samples is typically 16 kHz, we set the downsampling rates (DR) to 14 kHz, 12 kHz, and 10 kHz to evaluate the robustness of our method against downsampling. The results, shown in Table~\ref{tab:against_countermeasures}, indicate that as DR decreases, both the PSR and NISQA of our method show slight improvements. We suspect that this may be due to the removal of some useful high-frequency noise, leading to a slight improvement in audio quality and worse transcription results for adversarial examples. Overall, our method exhibits high robustness against downsampling.

\begin{table}[t]
\centering
\caption{Comparison of protection against temporal dependency detection with different $k$ settings.}
\label{tab:td_eval}
\resizebox{0.8\linewidth}{!}{
\begin{tabular}{c|c|c|c|c}
\Xhline{1px}
\multicolumn{2}{c|}{\textbf{}}& $k$ = 0.25 & $k$ = 0.50 & $k$ = 0.75 \\ \Xhline{1px}
\multicolumn{2}{c|}{Neekhara \etal~\cite{neekhara2019universal}}  & 50.63 & 76.04 & 88.29 \\ \hline
\multicolumn{2}{c|}{Zong \etal~\cite{zong2021targeted}}  & 50.70 & 80.50 & 94.94 \\ \hline
\multicolumn{2}{c|}{AdvDDoS~\cite{ge2023advddos}} & 50.65 & 76.41 & 88.26 \\ \hline
\multirow{3}{*}{\UAPSystemName} & $\tau$ = 0.5 & \textbf{49.00} & 74.15 & 88.16 \\ \cline{2-5}
 & $\tau$ = 0.4 & 50.47 & 69.76 & 82.35 \\ \cline{2-5}
 & $\tau$ = 0.3 & 50.40 & \textbf{65.14} & \textbf{74.47} \\ \Xhline{1px}
\end{tabular}
}
\vspace{-5mm}
\end{table}

\noindent\textbf{Temporal Dependency.}
It has been demonstrated that adversarial examples can disrupt the temporal dependency of the audio, which is a property that can be exploited to detect adversarial examples~\cite{yang2018characterizing}. 
Specifically, this method compares the similarity between the first $k$ portion transcription of an audio and that of the entire audio. If the similarity falls below a certain threshold, the example is classified as adversarial, where $k \in (0,1)$. 
To evaluate the robustness of \UAPSystemName against time-dependency-based detection, we set $k$ to 0.25, 0.50, and 0.75, and test the AUC of this countermeasure on three competitors and different variants of \UAPSystemName. 
The experimental results are shown in Table~\ref{tab:td_eval}, where a lower value indicates higher detection accuracy, meaning the protection is less resistant to temporal dependency detection. 
We find that in different $k$ settings, \UAPSystemName consistently achieves the lowest AUC, indicating the strongest resistance to this detection. Notably, with different values of $\tau$, the different variants of \UAPSystemName can achieve even lower AUCs, indicating better robustness. Moreover, according to the previous ablation study, when $\tau=0.4$, our method still outperforms competitors in terms of protection performance and audio quality. Therefore, our method allows users to select different variants based on their specific needs to achieve varying effects.
In summary, \UAPSystemName also exhibits excellent resilience to commonly used adaptive countermeasures, offering an additional layer of safeguarding for user privacy.

\noindent\textbf{Adaptive Countermeasures in Latent Space.}
An adversary with full knowledge of the operation of our system could attempt to remove the perturbation in the latent space using an autoencoder. Based on this, we design three adaptive countermeasures within the latent space. One straightforward approach involves using the autoencoder to reconstruct our adversarial example (marked as Recon), where the dimensionality reduction and expansion process could potentially mitigate or even eliminate the perturbation. Additionally, we experiment with local smoothing in the latent space during reconstruction (marked as LS-LS) or the addition of random noise (marked as LS-RN). However, the audio quality produced by all three methods is poor, with NISQA scores of 2.03, 2.05, and 1.92, indicating a significant deterioration compared to the unreconstructed examples. Additionally, we find that the reconstructed audio lost its usability, making it unintelligible to humans. Moreover, the WER for Recon, LS-LS, and LS-RN are 104.95\%, 101.32\%, and 105.96\%, respectively, demonstrating poor recognition performance by ASR systems. These results suggest that the aforementioned adaptive countermeasures are ineffective against our method, as they not only disrupt the adversarial perturbations but also degrade the audio usability. We hypothesize that this may stem from the encoder's inability to accurately encode our adversarial audio, implying that our adversarial examples exhibit strong robustness even against the model's encoding mechanism, further substantiating the transferability of \UAPSystemName.

\section{Conclusion}
In this paper, we propose a real-time privacy-preserving framework to avoid speech content leakage by unauthorized recognition, \UAPSystemName, whose core is transferable universal adversarial perturbations in latent space (\UAPName). By adding UAPs in the latent space and using the generative model to directly generate adversarial examples, we avoid introducing noise in the acoustic space and achieve better audio quality. Through target feature adaptation, we enhance the transferability of adversarial examples to unseen ASR models. 
Extensive experiments on 10 ASR models in over-the-line protections and over-the-air protections together 
demonstrate the superiority of both protection performance and audio quality, as well as practicality of \UAPSystemName. Further experiments show that \UAPSystemName can resist adaptive countermeasures. Our research offers protection of live user's speech content using adversarial examples, and provides sufficient evidence supporting its potential for widespread adoption in privacy-sensitive environments, \eg mass speech surveillance. 

\noindent\textbf{Limitations.}
Though \UAPSystemName has achieved good transferability across different target ASR models, the outputs of different target models for the same adversarial example are not always consistent. Even within the same model, the outputs may lack semantic coherence. This inconsistency and incoherence could potentially alert eavesdroppers~\cite{zeng2019multiversion}. In terms of consistency, it is important to note that none of the current methods studying transferable universal adversarial perturbations on ASR models can guarantee that all target models will produce the same output. Regarding coherence, we argue that maintaining semantic coherence in untargeted settings is an open and unexplored challenge in the field, which we consider as a direction for future work.

\section*{Acknowledgments}
We thank the reviewers and the shepherd for their constructive comments that significantly enhanced the paper. This research is supported in part by the National Natural Science Foundation of China under Grant No. U21B2020, the Beijing Natural Science Foundation under Grant No. QY24206, the Fundamental Research Funds for the Central Universities under Grant No. 2024ZCJH05 and the National Natural Science Foundation of China under Grant No. 62202064. This research is also supported in part by National University of Singapore. Jie Hao and Jin Song Dong are the corresponding authors of this paper.

\section*{Ethics Considerations}
\subsection*{1. Prevention of Misuse}
When AudioShield is applied in real-world scenarios, it may be unintentionally or maliciously misused. Specifically, we consider the following three potential misuse scenarios:

\noindent\textbf{Input Side of AudioShield.} This occurs primarily when the microphone on the user side (the sender) captures the user speech in a public environment. In this case, there is a possibility that the speech of others, without their consent, could be inadvertently recorded and then transmitted after being processed by AudioShield. To prevent this situation, we recommend implementing speech separation~\cite{pariente2020asteroid} and speaker verification~\cite{desplanques2020ecapa} during the deployment of AudioShield. This ensures that only registered users’ voices are input into AudioShield, thereby preventing unintended processing of speech by non-consenting individuals.
    
\noindent\textbf{Output Side of AudioShield.} This scenario involves unintentional misuse on the receiver side. The receiver, in a public setting, plays the protected speech received remotely, and the playback of this speech, being adversarial examples, may interfere with ASR systems in the public environment or accessibility tools used by others. However, we argue that even normal voice playback can cause disturbances in a public setting (since it may be perceived as noise by others). We recommend that users employ headphones or similar private playback methods in public settings to minimize sound leakage into the surrounding environment.
    
\noindent\textbf{Malicious Individuals.} A malicious individual can acquire adversarial audio generated by AudioShield (\eg by recording it at the receiver end). In the case of malicious misuse, this adversarial audio could affect the ASR applications of others in a shared space, leading to errors in speech recognition. However, since the adversarial examples generated by AudioShield are in an untargeted setting, they cannot be directed to cause specific malicious instructions in the target ASR. Therefore, the potential harm is not severe. That said, to minimize misuse, we require users to request permission and fill out the appropriate terms form, pledging not to misuse AudioShield. Only after our approval and authorization can they legally use AudioShield.

\subsection*{2. Responsible Disclosure}

Our adversarial testing aligns with the goals of improving security, privacy, and system robustness, which are often implicitly supported by API terms of service under the doctrine of fair use in research contexts. As long as the testing is non-disruptive, adheres to usage limits, and does not explicitly violate any clauses, it can be considered compliant. Furthermore, ethical research practices and the broader public interest in advancing privacy protection strengthen its justification under fair use principles. 
This demonstrates our commitment to transparency, legal adherence, and ethical norms. 

Since our research crafts adversarial examples on several commercial APIs, apart from its protective function, it also highlights the lack of robustness in these models. Therefore, we reported the vulnerability discovered to all service providers, including Google, Amazon, iFlytek, and Alibaba, through formal email correspondences. In our report, we also meticulously detailed the methodology employed in our method, with some demo audios generated by our method attached. We also outlined the potential risks that adversarial examples might trigger, as well as potential countermeasures. As our method serves as a protection for user privacy, we suggested that these vendors take it into consideration when addressing the identified security issue and making further improvements to their systems. We received responses with gratitude for our research and disclosure, acknowledging the value of our contributions to their ongoing efforts.

\subsection*{3. Experiment Ethics}
\noindent\textbf{Terms of Service and Permissions.}
Before conducting tests through commercial ASR APIs, we thoroughly reviewed and abided by the terms of service agreements to ensure proper use of the resources. Our experiments strictly adhere to these terms of service and privacy policies. Since the providers do not collect our experimental data, our adversarial testing does not impact other individuals or the commercial models themselves. 
Additionally, to reaffirm our commitment to ethical and legal standards in our experimental procedures, we proactively reached out to API providers, seeking explicit to use their APIs solely for academic research purposes to the best of our capability, and we received confirmation from Google and Amazon.

\noindent\textbf{User Study.}
The Human Research Ethics Committee of the authors’ affiliation determined that the study was exempt from further human subjects review. In our survey, all participants we recruited consented that their responses be used only for academic research. We did not collect any personal information that is unnecessary for our research.

\section*{Open Science}
All relevant source code and supporting scripts are made available at: \url{https://doi.org/10.5281/zenodo.14711220}. The repository includes a detailed README file with setup instructions, datasets, models and usage guidelines. All datasets we use are public datasets and we do not collect any additional data. To showcase the usability of the proposed method, an anonymous demo page with several audio examples is also included in this repository. The source code is released under the MIT License to ensure accessibility and fair use. Overall, we have made our artifact publicly available to facilitate reproducibility and foster further research.

\small{
\bibliographystyle{plain}
\bibliography{ref}
}

\appendix
\section{Proof}\label{append:proof}
\noindent\textbf{Proof of Theorem~\ref{probability_bound}.}
Since $z_1$ and $z_2$ can be seen as two independent random variables that satisfy the distribution of $\mathcal{E}\left(z|x\right)$, and $x \in \mathcal{X}$, where $\mathcal{X}$ denotes the input audio space, then $\mathcal{D}\left( {{z_1}} \right)$ and $\mathcal{D}\left( {{z_2}} \right)$ are also random variables.
Since $\mathcal{D}$ is $a$-Lipschitz, for any $z_1$ and $z_2$ in the latent space,
\begin{equation}
\begin{array}{l}
{\left\| {g\left( {{z_1}} \right) - g\left( {{z_2}} \right)} \right\|_\infty } \le a{\left\| {{z_1} - {z_2}} \right\|_\infty }\\
 \Leftrightarrow \mathcal{P}\left[ {{{\left\| {g\left( {{z_1}} \right) - g\left( {{z_2}} \right)} \right\|}_\infty } \le r} \right] \ge \mathcal{P}\left[ {a{{\left\| {{z_1} - {z_2}} \right\|}_\infty } \le r} \right]\\
 \Leftrightarrow \mathcal{P}\left[ {{{\left\| {g\left( {{z_1}} \right) - g\left( {{z_2}} \right)} \right\|}_\infty } \le r} \right] \ge 1 - \mathcal{P}\left[ {a{{\left\| {{z_1} - {z_2}} \right\|}_\infty } \ge r} \right].
\end{array}
\end{equation}

Due to Markov's Inequality, we have
\begin{equation}
\mathcal{P}\left[ {a{{\left\| {{z_1} - {z_2}} \right\|}_\infty } \ge r} \right] \le \frac{{{a^2}\mathcal{E}\left[ {{{\left\| {{z_1} - {z_2}} \right\|}_\infty }} \right]}}{{{r^2}}}.
\end{equation}

Though directly calculating the expectation for ${{\left\| {{z_1} - {z_2}} \right\|}_\infty }$ is difficult, the perturbation threshold $\tau $ helps us to obtain the following bound.
\begin{equation}
\begin{aligned}
\mathcal{P}\left[ {{{\left\| {g\left( {{z_1}} \right) - g\left( {{z_2}} \right)} \right\|}_\infty } \le r} \right] 
\ge & 1 - \mathcal{P}\left[ {a{{\left\| {{z_1} - {z_2}} \right\|}_\infty } \ge r} \right] \\
\ge & 1 - \frac{{{a^2}\mathcal{E}\left[ {{{\left\| {{z_1} - {z_2}} \right\|}_\infty }} \right]}}{{{r^2}}} \\
\ge & 1 - \frac{{{a^2}\tau }}{{{r^2}}}.
\end{aligned}
\end{equation}

To ensure a well-defined probability with non-negativity, the bound is set as $1 - \min \left\{ {1,\frac{{{a^2}\tau }}{{{r^2}}}} \right\}$. Therefore, the proof ends here.

\section{More Experimental Details and Analyses}

\begin{figure}[ht]
    \centering
    \includegraphics[width=0.98\linewidth]{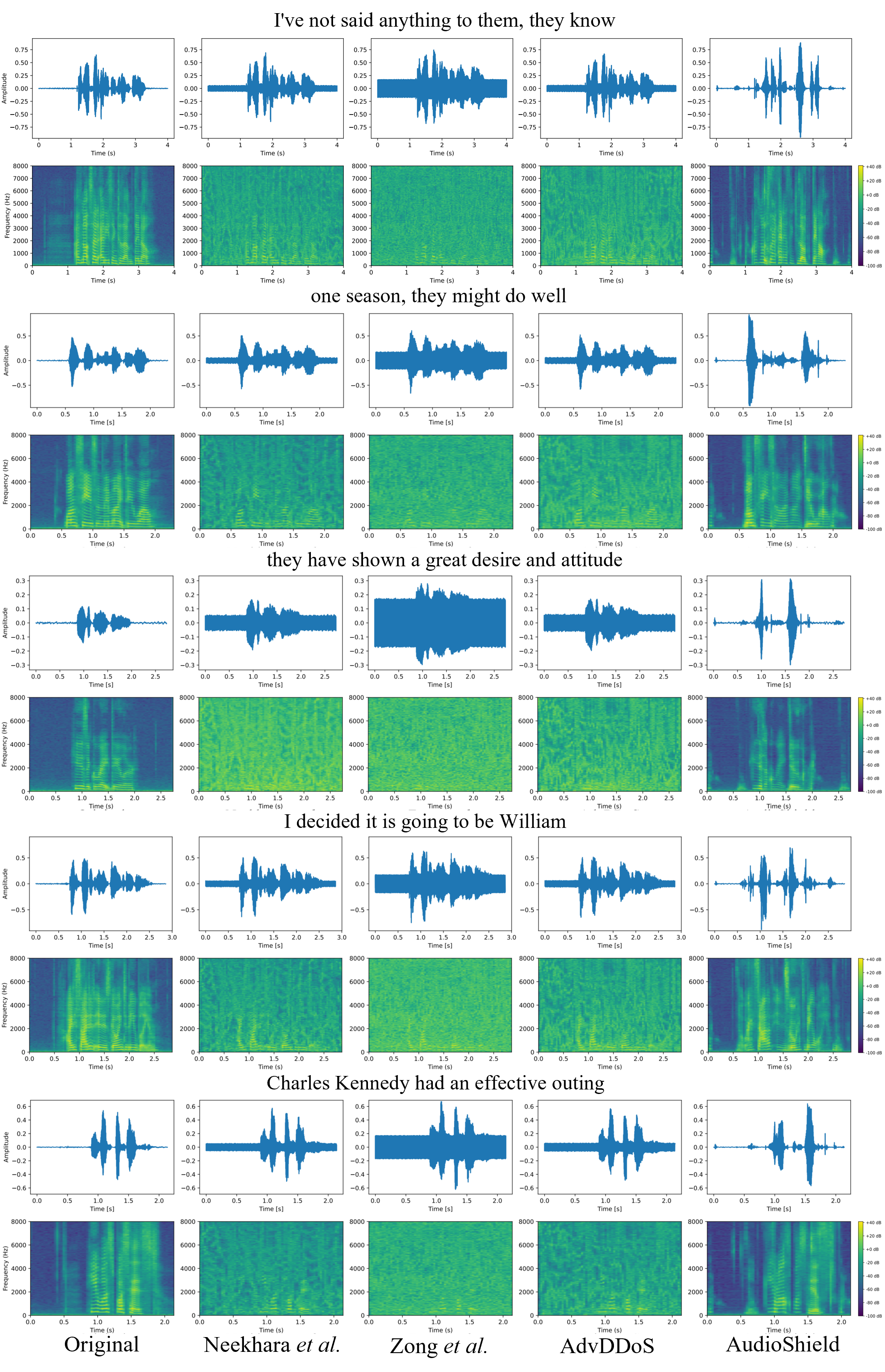}
    \caption{Visualization of the waveform and spectrogram corresponding to the text.}
    \label{fig:wav}
\end{figure}

\subsection{Visualization of Some Examples}
\label{append:vision}
Figure~\ref{fig:wav} shows the original waveform and spectrogram of five audio clips mentioned in Table~\ref{tab:objective}, along with those of their corresponding adversarial examples generated by the four methods. For our method, the generated adversarial examples exhibit significant changes in waveform shape compared to the original waveform after passing through the VAE. However, they are still close to the natural audio data distribution, resulting in higher quality. This illustrates the distinction between our method and traditional $\ell_p$-norm based methods, highlighting the ``unrestricted'' nature of \UAPSystemName. Traditional methods add perturbations directly in the original audio space and constrain them within a large $\ell_p$-norm range, limiting significant changes to the audio waveform shape.

For competitors, these waveform and spectrogram images vividly demonstrate the excessive noise introduced by their approaches. In the waveform images, the signals from the three competitors appear rougher, with more minor vibrations, indicating a higher noise content in these audios. In the spectrogram images, focusing on the time-frequency distribution of the signal, the energy distribution in the middle three spectrograms is more uniform, with less variation in frequency distribution, indicating more background noise in the signals. For changes in time-frequency features, the spectrograms of the original audio and the adversarial examples generated by our method exhibit clear variations in frequency components, indicating the presence of prominent speech events. In contrast, the spectrograms from the three competitors appear to be more stable with less noticeable changes, suggesting that their acoustic events are less prominent.

\subsection{Complete Result for Evaluation on Commercial ASR APIs}
\label{append:complete}
Our protection strategy involves training \UAPName locally in a targeted manner and then transferring it to the black-box ASR in an untargeted manner. Therefore, the selection of target texts during local training is crucial. According to our threat model, our objective is to generate a transferable universal adversarial perturbation. It is not necessary to ensure high protection performance under every target text; rather, we only need to identify a target text that achieves a good balance between high protection performance and audio quality for training. On the other hand, different target texts may result in variations in protection performance. Therefore, in order to better analyze the applicability of protection methods to different target texts, we conduct extensive evaluations using 10 common commands as target texts. 

Table~\ref{apd:complete_protection} presents the complete results of three competitors and \UAPSystemName on four target ASR APIs. Neekhara \etal \space is trained in an untargeted manner, resulting in only one result in the table. For the other three methods, targeted manipulations are performed locally with different target texts. The results in the table clearly demonstrate the superiority of our method, as it consistently achieves the highest protection performance across the majority of target texts. Moreover, for all target texts, our method also delivers the highest audio quality. Specifically, our method's NISQA score never falls below 2.00, regardless of the target text, whereas competitors never achieve a score above 2.00. The highest quality among competitors is shown by AdvDDoS under the target text ``play music'', with a score of 1.68, but even in this case, our method still outperforms it by 0.34.

Experimental results also show that the choice of target text influences protection performance, with both our method and competitors exhibiting significant variability across different texts. On average, our method achieves the highest PSR and NISQA, demonstrating strong transferability across all four target ASR models, as supported by the results in Table~\ref{tab:objective}. Considering both protection performance and audio quality, we select ``open the door'', which ranks in the top three for both metrics among the 10 texts, as the target text in our main experiments and make a fair comparison with competitors.

\subsection{Number of Recognition Failures}\label{append:number_of_NA}

\begin{table}[t]
\centering
\caption{Number of failed recognition examples.}
\label{tab:failure}
\resizebox{0.95\linewidth}{!}{
\begin{tabular}{c|c|c|c|c|c|c|c|c}
\Xhline{1px}
 \textbf{Method} & \textbf{Google} & \textbf{Amazon} & \textbf{iFlytek} & \textbf{Alibaba} & \textbf{Qwen-Audio} & \textbf{MooER} & \textbf{Whisper} & \textbf{Average} \\ 
\Xhline{1px}
Neekhara \etal & 315 & 1 & 0 & 4 & 79 & 1220 & 153 & \textbf{253.14} \\ \hline
Zong \etal & 1480.5 & 722.2 & 794.5 & 179.9 & 42 & 1472 & 1 & \textbf{670.30} \\ \hline
AdvDDoS & 458.8 & 17.3 & 9.7 & 21.6 & 71 & 1153 & 17 & \textbf{249.77} \\ \hline
\UAPSystemName & 469.1 & 21.9 & 12.4 & 15.4 & 82 & 366 & 23 & \textbf{141.40} \\ 
\Xhline{1px}
\end{tabular}
}
\vspace{-3mm}
\end{table}

Table~\ref{tab:failure} reports the number of examples with recognition failure in the tests. Among them, the recognition failure counts for the three methods, Zong \etal, AdvDDoS, and \UAPSystemName, on four commercial APIs are the averages of the failure counts, while the others are based on individual data groups. Each group contains a total of 2,000 test examples. Due to differences in the operational mechanisms and recognition capabilities of the models, the number of examples with recognition failure varies significantly across the seven models. We suspect that the primary reasons for recognition failure are the possible detection mechanisms within the commercial models and the low audio quality. On average, \UAPSystemName exhibits the smallest number of recognition failures, with only 141.40. This indirectly validates that the audio quality generated by our method is superior to that of the competitors.

\subsection{Questions of User Study}\label{append:user_study}
At the beginning, we informed each participant that all their responses were used solely for academic research, and we did not collect any of their personal information. We then clearly explained the task to the participants: ``You will now listen to audio clips and answer the corresponding questions.'' Specifically, for each audio clip, our instructions were as follows:
\begin{itemize}
    \item Please rate the quality of the following audio from 1 to 5, considering both the magnitude of the noise and the naturalness of the audio. The meaning of each score is given as follows:
    \begin{enumerate}
        \item Very loud noise, very poor naturalness.
        \item Noticeable noise, poor naturalness.
        \item Slight noise, average naturalness.
        \item Almost no noise, high naturalness.
        \item No noise at all, very high naturalness.
    \end{enumerate}
    \item Please transcribe the audio according to what you hear.
\end{itemize}

\begin{table}[t]
\centering
\caption{Complete results of protection performance on commercial ASR APIs.}
\label{apd:complete_protection}
\resizebox{1.0\linewidth}{!}{
\begin{tabular}{c|c|c|c|c|c|c|c|c|c|c}
\Xhline{1px}
\multirow{2}{*}{\textbf{Method}} & \multicolumn{2}{c|}{\textbf{Google}} & \multicolumn{2}{c|}{\textbf{Amazon}} & \multicolumn{2}{c|}{\textbf{iFlytek}} & \multicolumn{2}{c|}{\textbf{Alibaba}} & \multirow{2}{*}{\textbf{NISQA}} & \multirow{2}{*}{\textbf{Command}} \\ \cline{2-9}
 & \textbf{PSR} & \textbf{WER} & \textbf{PSR} & \textbf{WER} & \textbf{PSR} & \textbf{WER} & \textbf{PSR} & \textbf{WER} & \\ \Xhline{1px}
Neekhara \etal & 31.10 & 43.12 & 32.27 & 44.58 & 74.60 & 108.64 & 58.17 & 95.38 & 1.71 & - \\ \hline\hline
Zong \etal & 81.02 & 76.83 & 80.41 & 76.72 & 73.05 & 78.23 & 72.84 & 82.04 & 1.11 & \multirow{3}{*}{call my wife} \\ \cline{1-10}
AdvDDoS & 40.85 & 50.61 & 29.64 & 39.55 & 58.38 & 72.34 & 55.27 & 74.36 & 1.52 & \\ \cline{1-10}
\UAPSystemName & 92.27 & 88.77 & 94.01 & 91.01 & 80.84 & 88.91 & 88.63 & 93.03 & 2.42 & \\ \Xhline{1px}

Zong \etal & 71.39 & 70.60 & 61.07 & 63.12 & 56.37 & 67.24 & 62.22 & 73.31 & 1.25 & \multirow{3}{*}{make it warmer} \\ \cline{1-10}
AdvDDoS & 31.38 & 44.37 & 18.90 & 31.22 & 62.01 & 78.45 & 57.20 & 77.99 & 1.53 & \\ \cline{1-10}
\UAPSystemName & 88.84 & 87.96 & 81.89 & 81.60 & 72.55 & 81.38 & 77.47 & 85.37 & 2.11 & \\ \Xhline{1px}

Zong \etal & 58.31 & 64.14 & 42.65 & 52.45 & 47.42 & 61.41 & 61.04 & 74.69 & 1.00 & \multirow{3}{*}{navigate to my home} \\ \cline{1-10}
AdvDDoS & 36.31 & 47.72 & 24.57 & 35.47 & 64.61 & 74.54 & 62.93 & 83.92 & 1.63 & \\ \cline{1-10}
\UAPSystemName & 87.02 & 84.68 & 84.53 & 82.86 & 68.26 & 77.93 & 68.79 & 80.51 & 2.31 & \\ \Xhline{1px}

Zong \etal & 62.67 & 65.67 & 50.31 & 56.06 & 50.34 & 63.43 & 51.10 & 66.91 & 1.33 & \multirow{3}{*}{open the door} \\ \cline{1-10}
AdvDDoS & 31.02 & 43.82 & 19.19 & 30.80 & 39.85 & 55.28 & 63.76 & 101.11 & 1.54 & \\ \cline{1-10}
\UAPSystemName & 90.55 & 90.06 & 77.75 & 82.54 & 80.10 & 88.77 & 81.05 & 87.72 & 2.45 & \\ \Xhline{1px}

Zong \etal & 39.77 & 50.62 & 21.69 & 34.33 & 28.14 & 43.17 & 31.70 & 51.85 & 1.03 & \multirow{3}{*}{open the website} \\ \cline{1-10}
AdvDDoS & 34.09 & 44.50 & 19.24 & 30.23 & 55.75 & 69.71 & 35.65 & 53.86 & 1.53 & \\ \cline{1-10}
\UAPSystemName & 87.72 & 84.37 & 83.68 & 82.25 & 69.73 & 78.65 & 72.55 & 83.56 & 2.09 & \\ \Xhline{1px}

Zong \etal & 37.22 & 49.27 & 21.80 & 34.40 & 36.31 & 52.48 & 39.27 & 57.98 & 1.04 & \multirow{3}{*}{play music} \\ \cline{1-10}
AdvDDoS & 29.34 & 43.30 & 20.12 & 33.01 & 62.40 & 80.69 & 37.82 & 54.48 & 1.68 & \\ \cline{1-10}
\UAPSystemName & 77.06 & 78.02 & 79.42 & 79.09 & 86.75 & 69.16 & 54.40 & 68.62 & 2.02 & \\ \Xhline{1px}

Zong \etal & 83.16 & 77.88 & 83.12 & 79.42 & 72.23 & 78.04 & 73.05 & 81.38 & 1.08 & \multirow{3}{*}{send a text} \\ \cline{1-10}
AdvDDoS & 34.42 & 46.88 & 18.65 & 30.59 & 58.46 & 73.68 & 35.05 & 52.60 & 1.56 & \\ \cline{1-10}
\UAPSystemName & 81.81 & 82.28 & 85.60 & 84.14 & 66.58 & 78.39 & 57.92 & 72.62 & 2.36 & \\ \Xhline{1px}

Zong \etal & 52.40 & 59.33 & 35.96 & 46.33 & 40.95 & 55.08 & 49.24 & 66.19 & 1.34 & \multirow{3}{*}{take a picture} \\ \cline{1-10}
AdvDDoS & 46.41 & 53.14 & 42.38 & 49.66 & 81.20 & 95.06 & 70.81 & 106.40 & 1.37 & \\ \cline{1-10}
\UAPSystemName & 70.63 & 74.09 & 81.83 & 81.10 & 55.86 & 69.16 & 57.47 & 70.98 & 2.19 & \\ \Xhline{1px}

Zong \etal & 69.91 & 70.69 & 67.21 & 67.47 & 64.08 & 73.42 & 72.63 & 82.04 & 1.04 & \multirow{3}{*}{turn off the light} \\ \cline{1-10}
AdvDDoS & 33.37 & 44.96 & 25.35 & 35.97 & 57.43 & 73.83 & 59.67 & 86.04 & 1.32 & \\ \cline{1-10}
\UAPSystemName & 84.54 & 82.87 & 86.65 & 84.85 & 73.45 & 83.72 & 66.16 & 78.11 & 2.55 & \\ \Xhline{1px}

Zong \etal & 68.06 & 69.08 & 60.68 & 63.50 & 62.56 & 72.41 & 70.09 & 80.64 & 1.22 & \multirow{3}{*}{turn on airplane mode} \\ \cline{1-10}
AdvDDoS & 50.86 & 57.52 & 28.44 & 40.44 & 67.44 & 78.73 & 50.23 & 64.89 & 1.15 & \\ \cline{1-10}
\UAPSystemName & 91.32 & 88.35 & 91.05 & 87.73 & 81.85 & 88.72 & 82.75 & 92.00 & 2.33 & \\ \Xhline{1px}
\end{tabular}
}
\end{table}

\section{Discussion}
\noindent\textbf{Difference with Speaker Recognition Tasks.}
One prior work explored transferable universal adversarial perturbations on speaker recognition tasks~\cite{chen2023qfa2sr}. While this concept is not new in the audio domain, its application to ASR systems presents distinct challenges and complexities. Unlike classification tasks where the goal is to distinguish among a finite set of speakers, ASR systems are designed to convert continuous audio streams into text sequences, which is a more complex problem involving sequence-to-sequence mapping with a vast or even infinite output space. Furthermore, the transferability is more challenging due to the diversity of ASR architectures. However, speaker-specific features are easier to transfer between different speaker recognition systems. In addition, ASR requires contextual understanding, which involves processing contextual and syntactic information. Therefore, it is crucial to consider how perturbations influence both the immediate and broader context of speech when crafting perturbations.

%%%%%%%%%%%%%%%%%%%%%%%%%%%%%%%%%%%%%%%%%%%%%%%%%%%%%%%%%%%%%%%%%%%%%%%%%%%%%%%%
\end{document}